\pdfoutput=1
\PassOptionsToPackage{unicode}{hyperref}
\PassOptionsToPackage{hyphens}{url}
\PassOptionsToPackage{dvipsnames,svgnames,x11names}{xcolor}
\documentclass[
  a4paper,
  11pt,
  twoside]{predoc2cvdiag}
\usepackage{xcolor}
\usepackage{amsmath,amssymb}
\usepackage{iftex}
\ifPDFTeX
  \usepackage[T1]{fontenc}
  \usepackage[utf8]{inputenc}
  \usepackage{textcomp} 
\else 
  \usepackage{unicode-math} 
  \defaultfontfeatures{Scale=MatchLowercase}
  \defaultfontfeatures[\rmfamily]{Ligatures=TeX,Scale=1}
\fi
\usepackage{lmodern}
\ifPDFTeX\else
\fi
\IfFileExists{upquote.sty}{\usepackage{upquote}}{}
\IfFileExists{microtype.sty}{
  \usepackage[]{microtype}
  \UseMicrotypeSet[protrusion]{basicmath} 
}{}
\makeatletter
\@ifundefined{KOMAClassName}{
  \IfFileExists{parskip.sty}{%
    \usepackage{parskip}
  }{
    \setlength{\parindent}{0pt}
    \setlength{\parskip}{6pt plus 2pt minus 1pt}}
}{
  \KOMAoptions{parskip=half}}
\makeatother
\makeatletter
\ifx\paragraph\undefined\else
  \let\oldparagraph\paragraph
  \renewcommand{\paragraph}{
    \@ifstar
      \xxxParagraphStar
      \xxxParagraphNoStar
  }
  \newcommand{\xxxParagraphStar}[1]{\oldparagraph*{#1}\mbox{}}
  \newcommand{\xxxParagraphNoStar}[1]{\oldparagraph{#1}\mbox{}}
\fi
\ifx\subparagraph\undefined\else
  \let\oldsubparagraph\subparagraph
  \renewcommand{\subparagraph}{
    \@ifstar
      \xxxSubParagraphStar
      \xxxSubParagraphNoStar
  }
  \newcommand{\xxxSubParagraphStar}[1]{\oldsubparagraph*{#1}\mbox{}}
  \newcommand{\xxxSubParagraphNoStar}[1]{\oldsubparagraph{#1}\mbox{}}
\fi
\makeatother

\usepackage{longtable,booktabs,array}
\usepackage{calc} 
\usepackage{etoolbox}
\makeatletter
\patchcmd\longtable{\par}{\if@noskipsec\mbox{}\fi\par}{}{}
\makeatother
\IfFileExists{footnotehyper.sty}{\usepackage{footnotehyper}}{\usepackage{footnote}}
\makesavenoteenv{longtable}
\usepackage{graphicx}
\makeatletter
\newsavebox\pandoc@box
\newcommand*\pandocbounded[1]{
  \sbox\pandoc@box{#1}%
  \Gscale@div\@tempa{\textheight}{\dimexpr\ht\pandoc@box+\dp\pandoc@box\relax}%
  \Gscale@div\@tempb{\linewidth}{\wd\pandoc@box}%
  \ifdim\@tempb\p@<\@tempa\p@\let\@tempa\@tempb\fi
  \ifdim\@tempa\p@<\p@\scalebox{\@tempa}{\usebox\pandoc@box}%
  \else\usebox{\pandoc@box}%
  \fi%
}
\def\fps@figure{htbp}
\makeatother

\providecommand{\tightlist}{%
  \setlength{\itemsep}{0pt}\setlength{\parskip}{0pt}}

\usepackage[style=authoryear-comp,indexing=cite]{biblatex}
\runauthor{Vehtari and Bockting}
\runtitle{When can we trust Bayesian cross-validation computation?}
\renewcommand{\and}{\unskip{} and }
\renewbibmacro{in:}{}
\DeclareNameAlias{byeditor}{labelname}

\DeclareNameAlias{sortname}{given-family}
\DeclareNameAlias{default}{given-family}

\DeclareFieldFormat[article,inbook,incollection,inproceedings,thesis,unpublished,misc,online,manual,report]{title}{#1}

\renewbibmacro*{journal+issuetitle}{%
  \usebibmacro{journal}%
  \setunit*{\addcomma\space}%
  \printfield{volume}%
  \iffieldundef{number}{}{\printtext[parens]{\printfield{number}}}%
  \iffieldundef{eid}{}{\printtext{\addcolon}\printfield{eid}}%
  \newunit}

\DeclareFieldFormat{pages}{\ifentrytype{article}{#1}{pages~#1}}

\renewbibmacro*{date+extradate}{}
\renewbibmacro*{doi+eprint+url}{%
  \setunit{\addcomma\space}%
  \printlabeldateextra
  \newunit\newblock
  \iftoggle{bbx:doi}{\printfield{doi}}{}%
  \newunit\newblock
  \iftoggle{bbx:eprint}{\usebibmacro{eprint}}{}%
  \newunit\newblock
  \iftoggle{bbx:url}{\usebibmacro{url+urldate}}{}}
\renewbibmacro*{url+urldate}{\printfield{url}}

\AtEveryBibitem{\clearfield{issn}}

\makeatletter
\@ifpackageloaded{caption}{}{\usepackage{caption}}
\AtBeginDocument{%
\ifdefined\contentsname
  \renewcommand*\contentsname{Table of contents}
\else
  \newcommand\contentsname{Table of contents}
\fi
\ifdefined\listfigurename
  \renewcommand*\listfigurename{List of Figures}
\else
  \newcommand\listfigurename{List of Figures}
\fi
\ifdefined\listtablename
  \renewcommand*\listtablename{List of Tables}
\else
  \newcommand\listtablename{List of Tables}
\fi
\ifdefined\figurename
  \renewcommand*\figurename{Figure}
\else
  \newcommand\figurename{Figure}
\fi
\ifdefined\tablename
  \renewcommand*\tablename{Table}
\else
  \newcommand\tablename{Table}
\fi
}
\@ifpackageloaded{float}{}{\usepackage{float}}
\floatstyle{ruled}
\@ifundefined{c@chapter}{\newfloat{codelisting}{h}{lop}}{\newfloat{codelisting}{h}{lop}[chapter]}
\floatname{codelisting}{Listing}

\makeatother
\makeatletter
\@ifpackageloaded{caption}{}{\usepackage{caption}}
\@ifpackageloaded{subcaption}{}{\usepackage{subcaption}}
\makeatother
\usepackage{bookmark}
\IfFileExists{xurl.sty}{\usepackage{xurl}}{} 
\hypersetup{
  pdftitle={When can we trust Bayesian cross-validation computation?},
  pdfauthor={Aki Vehtari; Florence Bockting},
  colorlinks=true,
  linkcolor={blue},
  filecolor={Maroon},
  citecolor={Blue},
  urlcolor={Blue},
  pdfcreator={LaTeX via pandoc}}

\title{When can we trust Bayesian cross-validation computation?}
\usepackage{etoolbox}
\makeatletter
\providecommand{\subtitle}[1]{
  \apptocmd{\@title}{\par {\large #1 \par}}{}{}
}
\makeatother
\subtitle{Diagnostics and new methods}
\author{Aki Vehtari \and Florence Bockting}
\date{2026-11-30}
\begin{document}
\maketitle
\begin{abstract}
Cross-validation (CV) for Bayesian models is computed from a finite
number of posterior draws, so every CV estimate has a Monte Carlo error.
Current diagnostics check each fold separately, and brute-force, which
refits the model once for every fold, usually serves as the reference.
We give every CV method a self-diagnostic: an estimate of the error of
the total CV estimate, with a bias predictor of known sign, and a trust
rule that decides whether the estimate can be trusted. In 100 runs on
each of 37 combinations of models and data sets, we check whether these
self-diagnostics are calibrated. Brute-force can fail in specific
scenarios, but its self-diagnostic flags these cases. We introduce the
CV bridge, which combines the brute-force draws with the full-data draws
and provides a more accurate reference. We also introduce balanced
mixture importance sampling and, based on our results, propose a
practical workflow that tells a user what to do when a self-diagnostic
flags a problem.
\end{abstract}

\section{Motivation}\label{sec-intro}

After fitting a Bayesian model we usually want to know how well it
predicts new data, either to evaluate the model on its own or to choose
between several models \autocite{Vehtari+etal:PSIS-LOO:2017}. New data,
however, are not always available, and \emph{cross-validation} (CV)
therefore evaluates the predictive performance of a model with the
training data: it leaves out part of the data, the \emph{fold}, fits the
model to the rest, and evaluates the predictive density of the fold.

The direct way to compute CV, \emph{brute-force}, refits the model once
for every fold. It is often considered the ``gold standard'', but it is
also computationally expensive. \emph{Importance sampling} avoids the
refits by reweighting the draws from the full-data posterior so that
they represent the posterior without the fold
\autocite{Gelfand+Dey+Chang:1992,Vehtari+etal:PSIS-LOO:2017}. These
weights, however, can be unstable and can even have infinite variance
\autocite{Peruggia:1997,Epifani+MacEachern+Peruggia:2008}. \emph{Pareto
smoothed importance sampling} (PSIS) stabilizes them and provides a
diagnostic, \emph{Pareto-\(\hat{k}\)}, that warns when the reweighting
for a fold is not reliable \autocite{Vehtari+etal:PSIS:2024}.

Diagnostics are crucial for the practical use of CV methods, because
without them a user cannot know whether to trust a CV estimate. Current
practice relies mainly on fold-wise diagnostics, such as
Pareto-\(\hat{k}\), which checks the reliability of each fold.
Diagnostics that check the error of the total CV estimate itself (in the
following, \emph{self-diagnostics}) have received less attention.
Developing them requires new error estimates and a reference value
against which to check whether these estimates are calibrated. In CV,
brute-force is usually used as this reference
\autocite{Vehtari+etal:2016:LOO_for_GLVM,Vehtari+etal:PSIS-LOO:2017,Riha-Jedhoff-Buerkner-Vehtari:2026},
but its estimate can itself be wrong.
\textcite{Paananen+etal:2021:implicit} showed that it can be biased, for
example when one observation is an outlier, and
\textcite{Riha-Jedhoff-Buerkner-Vehtari:2026} showed that brute-force
LOGO estimates can be unreliable. This raises two questions:

\begin{itemize}
\tightlist
\item
  \textbf{Q1.} Is there a reference that is more accurate than
  brute-force (Section~\ref{sec-reference-check})?
\item
  \textbf{Q2.} Is brute-force a good reference, and can a
  self-diagnostic flag the cases where its estimate is wrong
  (Section~\ref{sec-bruteforce})?
\end{itemize}

Once an accurate reference is available, we can re-evaluate the
\emph{fast CV methods}: PSIS (Section~\ref{sec-background-psis}), PSIS
with moment matching (PSIS + MM, Section~\ref{sec-background-mm}) and
mixture importance sampling (MixIS, Section~\ref{sec-background-mixis}).
We can also check whether their self-diagnostics flag their failures
better than their current diagnostics do.

\begin{itemize}
\tightlist
\item
  \textbf{Q3.} How accurate are the fast methods, and can their
  self-diagnostics flag when they are not (Sections \ref{sec-fast-psis},
  \ref{sec-fast-mm} and \ref{sec-fast-mixis})?
\end{itemize}

Beyond better diagnostics, we ask whether the CV methods themselves can
be improved. For example, the mixture of MixIS can be difficult to
sample from when the leave-fold-out posteriors differ strongly
(Section~\ref{sec-background-mixis}).

\begin{itemize}
\tightlist
\item
  \textbf{Q4.} Does balancing the mixture reduce the error of MixIS
  (Section~\ref{sec-balanced-results})?
\end{itemize}

To answer these questions, we give every CV method a self-diagnostic: an
estimate of the error of its total CV estimate, together with a
\emph{trust rule} that decides whether the estimate can be trusted. We
check whether these self-diagnostics are calibrated on 37 combinations
of models and data sets. We introduce a new method, the \emph{CV
bridge}, which gives a more accurate reference than brute-force, and we
show that brute-force can be wrong but that its self-diagnostic flags
these cases. Of the fast methods, PSIS can be trusted only when every
Pareto-\(\hat{k}\) is below 0.7, moment matching is useful as a check
rather than as an estimator, and MixIS is accurate in more of our
experiments than PSIS; their self-diagnostics flag most of their
failures. We also introduce a balanced variant of MixIS, which is more
accurate than the original MixIS in most of our cases. Finally, we
combine the results into a workflow that starts with the cheapest method
and says what to do when a self-diagnostic flags a problem.

The rest of the paper is organized as follows:
Section~\ref{sec-background} reviews the CV methods in use today and
their diagnostics. Section~\ref{sec-framework} develops the
self-diagnostic and the trust rule. Section~\ref{sec-methods-new}
introduces two new variants of CV methods, and
Section~\ref{sec-integration} shows how integrating out group-specific
parameters makes CV methods more accurate. Section~\ref{sec-experiments}
checks whether the self-diagnostic is calibrated, for the existing and
the new CV methods. Section~\ref{sec-workflow} turns the results into a
practical workflow, and Section~\ref{sec-discussion} discusses our
findings and their limitations.

\section{Background}\label{sec-background}

In this section, we review the CV methods in use today, together with
their diagnostics.

\subsection{The predictive performance measure:
elpd}\label{sec-background-elpd}

In this paper, we measure predictive performance using the
\emph{expected log predictive density} (elpd), that is, the log density
that the model assigns to a new observation, averaged over the
distribution of new data \autocite{Vehtari+etal:PSIS-LOO:2017}. Let
\(y = (y_1, \dots, y_n)\) be the data and \(\theta\) the parameters of
the model. Covariates, where a model has them, are omitted from the
notation for simplicity. Throughout the paper, the observations are
modelled as conditionally independent given \(\theta\). The joint model
for the data and the parameters is then
\(p(y, \theta) = p(\theta) \prod_{i=1}^{n} p(y_i \mid \theta)\), where
\(p(\theta)\) is the prior distribution of the parameters and
\(p(y_i \mid \theta)\) is the data model for \(y_i\). Conditioning on
the data \(y\) gives the posterior
\(p(\theta \mid y) \propto p(\theta) \prod_{i=1}^{n} p(y_i \mid \theta)\)
\autocite[e.g.,][]{BDA3}. When group-specific parameters are integrated
out (Section~\ref{sec-integration}), the groups, and not the
observations, are modelled as conditionally independent given the
remaining parameters.

In \emph{leave-one-out} CV (LOO), each fold \(i\) consists of a single
observation, and \(y_{-i}\) denotes the data with observation \(i\)
removed. The LOO posterior, or more generally the \emph{leave-fold-out
posterior}, is
\(p(\theta \mid y_{-i}) \propto p(\theta) \prod_{j \neq i} p(y_j \mid \theta)\).
The model then predicts the left-out observation with the
\emph{leave-fold-out predictive density} \[
p(y_i \mid y_{-i}) = \int p(y_i \mid \theta)\, p(\theta \mid y_{-i})\,
  d\theta ,
\] the average of the likelihood of \(y_i\) over the leave-fold-out
posterior. CV estimates the elpd by the sum of the logarithms of these
densities \autocite{Vehtari+etal:PSIS-LOO:2017}: \[
\mathrm{elpd} = \sum_{i=1}^{n} \mathrm{elpd}_i, \qquad
\mathrm{elpd}_i = \log p(y_i \mid y_{-i}) .
\] Following \textcite{Vehtari+etal:PSIS-LOO:2017}, we use the word elpd
for this sum, which they denote \(\mathrm{elpd}_{\mathrm{loo}}\). It is
the CV estimate of the expected log predictive density for \(n\) new
observations. In this paper, the target of every method is this sum, and
the errors that we study are errors of its computation. Each
\(\mathrm{elpd}_i\) is the \emph{fold term} of fold \(i\), and the
methods of this paper differ only in how they compute these terms.

In \emph{leave-one-group-out} CV (LOGO) a whole group of observations is
left out, and the model must then predict a group it has never seen. The
fold is a group \(g\) with data \(y_g\), for example all measurements of
one patient, and \(y_{-g}\) denotes the data without the group. In a
hierarchical model each group has its own \emph{group-specific
parameters} \(b_g\), such as a patient-specific intercept, and the other
parameters \(\phi\) are shared by all groups. When group \(g\) is left
out, the remaining data contain no information about \(b_g\) beyond what
the \emph{population distribution} \(p(b_g \mid \phi)\) says about
groups in general. The leave-fold-out predictive density of the group is
therefore \[
p(y_g \mid y_{-g}) = \int p(y_g \mid \phi)\, p(\phi \mid y_{-g})\, d\phi,
\qquad
p(y_g \mid \phi) = \int p(y_g \mid b_g, \phi)\, p(b_g \mid \phi)\, db_g ,
\] where \(p(y_g \mid \phi)\) is the data model of the group averaged
over its own parameters. LOGO in this form measures how well the model
predicts a new group \autocite{Merkle+Furr+Rabe-Hesketh:2019}. In the
following, we describe the CV methods for LOO; for LOGO, replace \(i\)
by \(g\). The term \(p(y_g \mid \theta)\) is then
\(p(y_g \mid b_g, \phi)\) when \(b_g\) is sampled as a part of
\(\theta\), and \(p(y_g \mid \phi)\) when \(b_g\) is integrated out
(Section~\ref{sec-integration}).

\subsection{Brute-force}\label{sec-background-bf}

In brute-force CV we obtain \(S\) draws \(\theta_1, \dots, \theta_S\)
from the leave-fold-out posterior, usually with a new run of
\emph{Markov chain Monte Carlo} (MCMC), and estimate
\(p(y_i \mid y_{-i})\) by \[
\hat{p}_i = \frac{1}{S} \sum_{s=1}^{S} p(y_i \mid \theta_s),
\qquad
\widehat{\mathrm{elpd}}_i = \log \hat{p}_i .
\] For any method, \(\hat{p}_i\) denotes its estimate of
\(p(y_i \mid y_{-i})\). The cost of brute-force is one MCMC run per
fold.

Like every CV method in this paper, brute-force computes the estimate
from a finite number of posterior draws, so the estimate has a
\emph{Monte Carlo error}: if the computation is repeated with a new
random seed, that is, in a new \emph{run}, the estimate changes. Because
every fold has its own MCMC run, the Monte Carlo errors of the fold
terms are independent of each other, so the variance of the estimated
elpd is the sum of the variances of the fold terms
\autocite{Cooper-Vehtari-Forbes-etal:2024}. For the methods that use the
same draws for all folds, this is not true, and
Section~\ref{sec-framework-diag} shows why the difference matters.

Brute-force is usually treated as exact, and in practice it is rarely
checked
\autocite{Vehtari+etal:2016:LOO_for_GLVM,Vehtari+etal:PSIS-LOO:2017}.
Its random error can be quantified by the \emph{Monte Carlo standard
error} (MCSE), an estimate of the standard deviation over runs, of each
fold term and of their sum \autocite{Cooper-Vehtari-Forbes-etal:2024},
but so far this has mainly been used to show that the error is small,
not as a diagnostic that decides whether an estimate can be trusted.

Brute-force, however, can itself be biased.
\textcite{Paananen+etal:2021:implicit} fitted a normal model to data
with one outlier, a case where the leave-one-out value is known
analytically, and found brute-force increasingly biased as the magnitude
of the outlier increased. They used Pareto-\(\hat{k}\) as a diagnostic
for brute-force, and it flagged these runs.
\textcite{Riha-Jedhoff-Buerkner-Vehtari:2026}, in contrast, observed
brute-force estimates that stayed close to their reference while
Pareto-\(\hat{k}\) exceeded the PSIS threshold 0.7, and concluded that
other diagnostics may be needed. Pareto-\(\hat{k}\) sees that the
observed values \(p(y_i \mid \theta_s)\) look like draws from a
heavy-tailed distribution, but if that tail ends at an unknown point, it
cannot tell how large the error is (Section~\ref{sec-diag-bf}).

\subsection{Importance sampling: PSIS}\label{sec-background-psis}

Importance sampling avoids the refits by using the draws from the
full-data posterior, which is then the common \emph{proposal}
distribution for all the leave-fold-out posteriors. The full-data
posterior is the leave-fold-out posterior multiplied by the likelihood
of the left-out observation, up to a constant, so dividing by
\(p(y_i \mid \theta_s)\) removes the contribution of fold \(i\). Each
draw therefore gets the weight
\(w_{i,s} \propto 1/p(y_i \mid \theta_s)\), and the estimate becomes
(after some algebra) \[
\hat{p}_i =
  \left( \frac{1}{S} \sum_{s=1}^{S} \frac{1}{p(y_i \mid \theta_s)}
  \right)^{-1},
\] a harmonic mean of the likelihood values. A harmonic mean is
dominated by its smallest terms: a few draws under which \(y_i\) is very
unlikely get very large weights, and the weights can even have infinite
variance
\autocite{Peruggia:1997,Epifani+MacEachern+Peruggia:2008,Silva+Zanella:2024:robustloo}.
Such draws are common when the full-data posterior and the
leave-fold-out posterior are very different, which is typical for highly
influential observations \autocite{Vehtari+etal:PSIS:2024}, and for the
same reason importance sampling often fails in LOGO, where a fold
contains many observations
\autocite{Vehtari-Lampinen:2002,Cooper-Vehtari-Forbes-etal:2024,Riha-Jedhoff-Buerkner-Vehtari:2026}.

Pareto smoothed importance sampling (PSIS) reduces the influence of the
largest weights \autocite{Vehtari+etal:PSIS:2024}. It fits a
\emph{generalized Pareto distribution} (GPD), a model for the upper tail
of a distribution, to the \(M = \min(0.2S, 3\sqrt{S})\) largest weights
and replaces these weights by smoothed values from the fitted
distribution. The shape parameter of the fitted GPD, Pareto-\(\hat{k}\),
measures how heavy the tail of the weights is, and PSIS uses it as its
diagnostic. With the usual number of draws (\(S > 2000\)), the PSIS
estimate of a fold term is reliable when Pareto-\(\hat{k} \leq 0.7\).
For folds with Pareto-\(\hat{k} > 0.7\),
\textcite{Vehtari+etal:PSIS-LOO:2017} recommend refitting. The loo
package also reports an MCSE for each fold term and combines them into
an MCSE of the total as if the fold errors were independent
\autocite{Vehtari+etal:2026:loo}, which ignores that all folds use the
same full-data posterior draws. Section~\ref{sec-framework-diag}
estimates the error of the total, including the correlation between
folds, and Section~\ref{sec-fast-psis} tests it.

\paragraph{Moment matching}\label{sec-background-mm}

\emph{Moment matching} (MM) tries to repair the importance sampling of a
fold without a refit \autocite{Paananen+etal:2021:implicit}. The
weighted mean of the draws estimates the mean of the leave-fold-out
posterior. MM shifts the draws so that their mean equals this weighted
mean; if that is not enough, it also rescales the draws so that their
variances match, and finally it matches their covariance. After each
step the weights are recomputed for the transformed draws, and
Pareto-\(\hat{k}\) decides whether to continue.
\textcite{Paananen+etal:2021:implicit} combine the steps into a stepwise
algorithm: use PSIS when Pareto-\(\hat{k}\) is small, use MM when it is
not, and refit when MM does not bring Pareto-\(\hat{k}\) below the
threshold. They regard a matched fold as reliable when the
Pareto-\(\hat{k}\) of its new weights is at most 0.7 and showed on a
simple example with a known leave-one-out value that the matched
estimate has negligible bias. However, in their more realistic examples
they reported only how often MM brings Pareto-\(\hat{k}\) below the
threshold.

More recent studies report mixed results on MM as a repair.
\textcite{Riha-Jedhoff-Buerkner-Vehtari:2026} found that MM reduced
Pareto-\(\hat{k}\) but did not always improve the estimate, and
suggested that the reuse of the same draws may be the reason.
\textcite{Bindoff:2026} found that MM reduced the error of PSIS only
slightly for hierarchical models. Whether MM should be used to repair
unstable weights therefore needs further study, and so does the question
of which alternatives exist when it should not be used.
Section~\ref{sec-fast-mm} answers the first question, and
Section~\ref{sec-methods-new} and Section~\ref{sec-integration} describe
two alternatives.

\subsection{Mixture importance sampling}\label{sec-background-mixis}

The extreme weights of PSIS arise when its proposal, the full-data
posterior, is far from the leave-fold-out posterior. \emph{Mixture
importance sampling} \autocite[MixIS,][]{Silva+Zanella:2024:robustloo}
uses a different proposal. It runs MCMC once on \[
q_{\mathrm{mix}}(\theta) \propto p(\theta \mid y)
  \sum_{i'=1}^{n} \frac{1}{p(y_{i'} \mid \theta)},
\] which is a mixture of all leave-fold-out posteriors, and it gives
draw \(s\) the weight \[
v_{i,s} = \frac{1 / p(y_i \mid \theta_s)}
               {\sum_{i'} 1 / p(y_{i'} \mid \theta_s)}
\] for fold \(i\). Each weight is the share of one fold in the sum, so
it is at most one, and the estimator has finite asymptotic variance
under weak conditions \autocite{Silva+Zanella:2024:robustloo}. The cost
is one additional MCMC run. The mixture gives more weight to folds with
a small \(p(y_i \mid y_{-i})\), which are the hardest to estimate, but
when the leave-fold-out posteriors are very different, the mixture can
have several modes and be difficult to sample
\autocite{Silva+Zanella:2024:robustloo}. The estimator stays consistent
for any positive weights of the mixture components, that is, factors in
front of the terms \(1/p(y_{i'} \mid \theta)\).
\textcite{Silva+Zanella:2024:robustloo} note that unequal weights might
improve efficiency, but they give no rule for choosing them and use
equal weights.

While the advantages of MixIS are well understood,
\textcite{Silva+Zanella:2024:robustloo} propose no MCSE or diagnostic
for it, so users have been uncertain when to trust its estimate.
Section~\ref{sec-diag-mixis} provides a self-diagnostic, the balanced
variant introduced in Section~\ref{sec-methods-new} makes MixIS more
robust, and Section~\ref{sec-fast-mixis} and
Section~\ref{sec-balanced-results} evaluate both.

\subsection{Bridge sampling}\label{sec-background-bridge}

The leave-fold-out predictive density can also be written as the ratio
of two \emph{marginal likelihoods}, \[
p(y_i \mid y_{-i}) = \frac{p(y)}{p(y_{-i})},
\] where \(p(y) = \int p(\theta)\, p(y \mid \theta)\, d\theta\) and
\(p(y_{-i}) = \int p(\theta)\, p(y_{-i} \mid \theta)\, d\theta\).
Marginal likelihoods are also the normalizing constants for the
posterior distributions. We can estimate normalizing constants with
bridge sampling \autocite{Meng+Wong:1996,Gronau+etal:2017}. For two
unnormalized distributions \(q_1\) and \(q_2\) with normalizing
constants \(Z_1\) and \(Z_2\), \[
\frac{Z_1}{Z_2} =
  \frac{\mathrm{E}_{2}\!\left[ q_1(\theta)\, h(\theta) \right]}
       {\mathrm{E}_{1}\!\left[ q_2(\theta)\, h(\theta) \right]}
\] for any function \(h\), the \emph{bridge function}, for which both
expectations are finite and the denominator is not zero;
\(\mathrm{E}_1\) and \(\mathrm{E}_2\) are expectations under the two
normalized distributions. If one of the two distributions has a known
normalizing constant, the identity gives the normalizing constant of the
other. A common choice is a normal distribution with mean and covariance
matched to the posterior draws of the target distribution. Importance
sampling is the special case \(h = 1/q_2\), which needs draws from only
one of the two distributions. \textcite{Meng+Wong:1996} give the bridge
function with the smallest asymptotic error, compute the estimate by a
simple iteration, and show that using draws from both distributions can
reduce the error by orders of magnitude when the two distributions are
far apart. With this bridge function, each term of the two sample
averages is bounded, so a few extreme draws cannot dominate the
estimate. The method still requires the two distributions to overlap
\autocite{Meng+Wong:1996,Gronau+etal:2017}. When the overlap is poor,
the draws can first be moved closer to each other, for example by a
shift or a rescaling \autocite{Meng+Wong:1996}.

\textcite{Micaletto+Vehtari:2025} derive an MCSE for bridge sampling
that accounts for the autocorrelation of MCMC draws, and show that it is
reliable when it is below 0.3. They also found that Pareto-\(\hat{k}\),
computed on the terms of the bridge estimator, is not a useful
diagnostic for bridge sampling.
\textcite{Riha-Jedhoff-Buerkner-Vehtari:2026} use bridge sampling with a
multivariate normal proposal distribution to estimate \(p(y)\) and
\(p(y_{-g})\) and so obtain a reference value for LOGO. They also
investigate the use of bridge sampling to estimate \(p(y_g \mid \phi)\)
at each posterior draw of \(\phi\) and use these estimates with PSIS
(see also Section~\ref{sec-integration}).

Section~\ref{sec-methods-new} introduces the \emph{CV bridge} method,
which uses the draws from both the full-data posterior and the
leave-fold-out posterior and applies the bridge iteration directly to
the ratio \(p(y)/p(y_{-i})\) without normal approximations.
Section~\ref{sec-reference-bridge} explains why this has a smaller error
than estimating the two marginal likelihoods separately, and
Section~\ref{sec-reference-check} shows that it is accurate enough to
serve as the reference in the experiments.

\section{Self-diagnostics}\label{sec-framework}

To answer the questions introduced in Section~\ref{sec-intro}, a method
must report how wrong its estimate is. This section derives such a
self-diagnostic in general and then computes it for brute-force, PSIS,
MM and MixIS.

\subsection{The error of an elpd estimate}\label{sec-framework-diag}

A method computes the elpd from the \(S\) posterior draws of one run, so
its estimate has a Monte Carlo error. For one run, method \(m\) returns
fold estimates \(\widehat{\mathrm{elpd}}_{m,i}\) and their sum
\(\widehat{\mathrm{elpd}}_m = \sum_i \widehat{\mathrm{elpd}}_{m,i}\).
For a specific method, its name replaces \(m\), as in
\(\widehat{\mathrm{elpd}}_{\mathrm{bf},i}\), and other quantities of the
method carry the name in the same place, as in
\(\mathrm{mcse}_{\mathrm{psis},i}\). Over repeated runs with the data
fixed, the mean squared error of the sum splits into two components, a
variance and a squared bias: \[
\mathrm{E}\Big[\Big(\sum_i \widehat{\mathrm{elpd}}_{m,i} - \mathrm{elpd}\Big)^2\Big]
  = \underbrace{\mathrm{Var}\Big(\sum_i \widehat{\mathrm{elpd}}_{m,i}\Big)}_{\mathrm{variance}} + \underbrace{\Big(\mathrm{E}\Big[\sum_i \widehat{\mathrm{elpd}}_{m,i}\Big] - \mathrm{elpd}\Big)^2}_{\mathrm{bias}^2}.
\] Both components are defined over many runs, but a user estimates the
elpd only once, from a single run. A self-diagnostic must therefore be
able to estimate both components from the draws of this single run.

To estimate the variance component, we use the \emph{squared MCSE},
\(\mathrm{mcse}_{\mathrm{total}}^2\), which splits into the fold
variances and their covariances:
\begin{equation}\phantomsection\label{eq-mcse-total}{
\mathrm{mcse}_{\mathrm{total}}^2
  = \sum_{i} \underbrace{\widehat{\mathrm{Var}}(\widehat{\mathrm{elpd}}_{m,i})}_{\mathrm{mcse}_i^2}
  + \sum_{i \neq i'} \widehat{\mathrm{Cov}}\big(\widehat{\mathrm{elpd}}_{m,i},
      \widehat{\mathrm{elpd}}_{m,i'}\big).
}\end{equation} To estimate the bias component, we introduce the
\emph{bias predictor} \(\mathrm{bias}_{\mathrm{pred}}\), which we derive
below (\ref{eq-biaspred}). Replacing the two components of the
decomposition by their estimates gives the \emph{predicted root mean
square error} (predicted RMSE), \[
\mathrm{rmse}_{\mathrm{pred}}^2
  = \mathrm{mcse}_{\mathrm{total}}^2 + \mathrm{bias}_{\mathrm{pred}}^2,
\] which is the \emph{error estimate} of the self-diagnostic. It is
computed from the same draws as the estimate and can therefore fail for
the same reasons \autocite{Vehtari-Lampinen:2002}. A self-diagnostic
therefore has two further parts: \emph{reliability checks}, which test
whether the error estimate can be believed, and a \emph{trust rule},
which combines the error estimate and the reliability checks into one
decision, trusted or not trusted.

In every method of this paper, the fold estimate is built from
logarithms of Monte Carlo means. We write such a mean as
\(\bar{a}_i = \frac{1}{S} \sum_s a_{i,s},\) where the values \(a_{i,s}\)
are computed from the draws \(\theta_s\). The fold estimate is
\(\log \bar{a}_i\) for brute-force, \(-\log \bar{a}_i\) for PSIS, and
the difference of two such logarithms for MixIS. The upcoming
subsections on the three methods define \(a_{i,s}\) for each. The
derivations that follow are stated for a single logarithm of a mean,
\(\log \bar{a}_i\).

\paragraph{Variance component}\label{variance-component}

Equation (\ref{eq-mcse-total}) builds \(\mathrm{mcse}_{\mathrm{total}}\)
from two kinds of terms, the fold-wise variances \(\mathrm{mcse}_i^2\)
and the covariances between folds. The covariance term depends on how
the draws are used. When every fold estimate is computed from draws of
its own, as in brute-force, the covariances are zero, and the total MCSE
is the \emph{root sum of squares} (RSS) of the fold MCSEs
\autocite{Cooper-Vehtari-Forbes-etal:2024}, \[
\mathrm{mcse}_{\mathrm{total}} = \mathrm{RSS}
  = \sqrt{\textstyle\sum_i \mathrm{mcse}_i^2} .
\] When all fold estimates are computed from the same draws, as in PSIS
and MixIS, a draw that increases one fold term tends to increase other
fold terms too. The fold errors are then positively correlated, and the
RSS would underestimate the total MCSE. To include the covariances, we
use the influence of each draw. The \emph{influence} \(\psi_{i,s}\) of
draw \(s\) on fold \(i\) is the first-order change in
\(\widehat{\mathrm{elpd}}_{m,i}\) when draw \(s\) receives slightly more
weight than the other draws, a construction known as the
\emph{infinitesimal jackknife} (IJ)
\autocite{Giordano-Stephenson-Liu-etal:2019}. For each draw, we sum its
influences on all folds and then square the sum, which includes the
covariance terms. For the logarithm of a mean, \[
\psi_{i,s} = \frac{a_{i,s} - \bar{a}_i}{S\,\bar{a}_i},
\] which is the empirical influence function divided by \(S\), so that
\(\sum_s \psi_{i,s}^2\) is the squared fold MCSE of (\ref{eq-delta})
with \(S\) in place of the ESS. The influence of draw \(s\) on the total
is \(\psi_s = \sum_i \psi_{i,s}\), and the total MCSE is
\begin{equation}\phantomsection\label{eq-ij}{
\mathrm{mcse}_{\mathrm{total}}^2
  = \frac{1}{\mathrm{relESS}(\psi)} \sum_s \psi_s^2
  = \frac{1}{\mathrm{relESS}(\psi)}
    \Bigg(\underbrace{\sum_i \sum_s \psi_{i,s}^2}_{\text{fold variances}}
    + \underbrace{\sum_{i \neq i'} \sum_s \psi_{i,s}\, \psi_{i',s}}_{\text{covariances}}
    \Bigg),
}\end{equation} where the expanded form shows the two kinds of terms of
(\ref{eq-mcse-total}), and \(\mathrm{relESS}(\psi)\) is the
\emph{relative effective sample size} of the \(\psi_s\): their
\emph{effective sample size} (ESS), the number of independent draws that
would give the same variance as the autocorrelated MCMC draws, divided
by \(S\). The fold terms of (\ref{eq-ij}) therefore differ from the
\(\mathrm{mcse}_i^2\) of (\ref{eq-delta}), which use the ESS of each
fold. The influences of PSIS and MixIS are given in their subsections
below.

The fold-wise variances follow from the \emph{delta method}
\autocite[Ch. 3]{vanderVaart:1998}, which converts the error of a mean
into the error of its logarithm. For the logarithm of a Monte Carlo
mean, it gives to first order
\begin{equation}\phantomsection\label{eq-delta}{
\mathrm{mcse}_i^2 = \frac{\widehat{\mathrm{Var}}(\bar{a}_i)}{\bar{a}_i^2}
  = \frac{\mathrm{sd}(a_{i,s})^2}{\mathrm{ESS}(a_i)\,\bar{a}_i^2},
}\end{equation} where the standard deviation and \(\mathrm{ESS}(a_i)\),
the ESS of the values \(a_{i,s}\) of fold \(i\), are taken over the
draws; the ESS replaces \(S\) because MCMC draws are autocorrelated.
Equation (\ref{eq-delta}) also holds for \(-\log \bar{a}_i\) (PSIS). For
\(\log \bar{a}_i - \log \bar{b}_i\) (MixIS), it applies to both means,
and their covariance enters as well; the MixIS subsection gives the
formula. Equation (\ref{eq-delta}) works when the values \(a_{i,s}\)
have similar magnitude across the draws. When a few draws give much
larger values \(a_{i,s}\) than all others, their distribution has a
heavy upper tail. For brute-force, where
\(a_{i,s} = p(y_i \mid \theta_s)\), this happens for example when
\(y_i\) is unlikely under most draws and very likely under a few. The
largest values are then rarely drawn, the sample standard deviation
misses them, and the MCSE is too small exactly when the error is large.

Instead of relying on the draws alone, we can model the tail. As PSIS
does for its weights, a GPD fitted to the largest values models the part
of the tail that the draws miss \autocite{Vehtari+etal:PSIS:2024}. When
its shape \(\hat{k} \geq 0.5\), however, the fitted tail has infinite
variance. To get a finite variance, we truncate the tail model at a
point beyond which no values are expected. The variance of the truncated
tail replaces the sample variance in (\ref{eq-delta}). We truncate for
every \(\hat{k} > 0\). The methods differ in how they use this truncated
variance. Brute-force uses it for the MCSE, and we take as truncation
point the smaller of two values: the largest plausible value of
\(a_{i,s}\), and the 95\% quantile of the largest of the \(M\) tail
values under the fitted GPD. PSIS uses it only for the bias predictor;
for its weights we have no largest plausible value, so the truncation
point is this quantile. MixIS does not use it: its fold weights are
bounded, and for its common weights the GPD fit gives only a
Pareto-\(\hat{k}\) warning. The method subsections below give the
details. Section~\ref{sec-diag-bf} defines the largest plausible value
and the truncated variance, and Appendix \ref{sec-pmax} compares the
choices for it.

\paragraph{Bias component}\label{bias-component}

Because the fold estimate is built from logarithms of Monte Carlo means,
it is biased even when the means themselves are not. The logarithm is
concave, so by Jensen's inequality
\begin{equation}\phantomsection\label{eq-jensen}{
\mathrm{E}\big[\log \bar{a}_i\big] \leq \log \mathrm{E}\big[\bar{a}_i\big],
}\end{equation} where the expectation is over runs. When
\(\log \mathrm{E}[\bar{a}_i]\) is the target of the fold estimate, the
logarithm of a mean is therefore biased downward, and minus the
logarithm of a mean upward, for every finite \(S\). Jensen's inequality
gives the sign of this bias but not its magnitude. A second-order Taylor
expansion of \(\log \bar{a}_i\) around \(\mathrm{E}[\bar{a}_i]\) gives
the magnitude, \[
\mathrm{E}\big[\log \bar{a}_i\big]
  \approx \log \mathrm{E}\big[\bar{a}_i\big]
  - \frac{\mathrm{Var}(\bar{a}_i)}{2\,\mathrm{E}[\bar{a}_i]^2},
\] and by (\ref{eq-delta}) the last term is half the squared fold MCSE.
The bias predictor of a fold is therefore
\begin{equation}\phantomsection\label{eq-biaspred}{
\mathrm{bias}_{\mathrm{pred},i} = \mp\tfrac{1}{2}\,\mathrm{mcse}_i^2
\quad \text{for } \pm \text{ the logarithm of a mean},
}\end{equation} and the bias of the elpd is the sum of these fold
biases. Now that we can predict not only the sign but also the magnitude
of the bias, it is tempting to correct the estimate by subtracting the
predicted bias. We do not do this, for three reasons. First, the
second-order term is accurate only when the bias is small. Second, under
a heavy tail it underestimates the bias, so the corrected estimate would
still be biased. Third, the subtraction adds the Monte Carlo error of
the predictor to the estimate. Section~\ref{sec-experiments} shows how
large the second effect can be. The sign of
\(\mathrm{bias}_{\mathrm{pred}}\), however, is useful even when its
magnitude is not. It enters the predicted RMSE, the trust rule, and an
\emph{asymmetric interval} for the error
\(\widehat{\mathrm{elpd}}_m - \mathrm{elpd}\) that is wider on the side
of \(\mathrm{bias}_{\mathrm{pred}}\), \[
\left[\, -1.96\,\mathrm{mcse}_{\mathrm{total}}
  + \min(0, \mathrm{bias}_{\mathrm{pred}}),\;
  1.96\,\mathrm{mcse}_{\mathrm{total}}
  + \max(0, \mathrm{bias}_{\mathrm{pred}}) \,\right] .
\] For the user, the sign tells on which side of the estimate the elpd
more likely lies: above it for a downward bias, below it for an upward
bias.

The following subsections give, for each method, the error estimate, the
reliability checks and the trust rule. Appendix \ref{sec-methods-table}
collects them in one table.

\subsection{Brute-force}\label{sec-diag-bf}

For brute-force, the fold estimate is \[
\widehat{\mathrm{elpd}}_{\mathrm{bf},i} = \log \hat{p}_i,
\qquad
\hat{p}_i = \frac{1}{S}\sum_s p(y_i \mid \theta_s),
\] with \(S\) draws from the leave-fold-out posterior of fold \(i\);
that is, \(a_{i,s} = p(y_i \mid \theta_s)\) and
\(\bar{a}_i = \hat{p}_i\). Every fold has its own MCMC run, so the
covariances in (\ref{eq-mcse-total}) vanish and the total MCSE is the
RSS of the fold MCSEs. The values \(p(y_i \mid \theta_s)\) can have a
heavy upper tail, so the fold MCSE uses the truncated tail model instead
of (\ref{eq-delta}).

We fit a GPD to the \(M\) largest values \(p(y_i \mid \theta_s)\), with
threshold \(c\) (the largest value below these \(M\)), scale
\(\hat{\sigma}\) and shape \(\hat{k}\)
\autocite{Vehtari+etal:PSIS:2024}. Under the fitted GPD, the
standardized tail value
\(X = 1 + \hat{k}\,(p(y_i \mid \theta_s) - c)/\hat{\sigma}\) follows a
Pareto distribution when \(\hat{k} > 0\). When \(\hat{k} \leq 0\), the
fitted GPD has a finite mean and variance, and they are used without the
truncation described next. The values \(p(y_i \mid \theta_s)\) may be
bounded, but when the bound lies beyond the observed draws, the draws
behave as if the tail were unbounded, and \(\hat{k}\) is positive
\autocite[Sec. 3.2.9]{Vehtari+etal:PSIS:2024}. The draws therefore do
not show where the tail ends, and we supply a \emph{largest plausible
value} \(p_{\max}\) of \(p(y_i \mid \theta_s)\), defined below. On the
standardized scale it becomes
\(X \leq \tau = 1 + \hat{k}\,(p_{\max} - c)/\hat{\sigma}.\) In addition,
the tail model describes only the \(M\) largest values, and the largest
of \(M\) draws from the fitted Pareto distribution is below \[
\tau_M = \big(1 - 0.95^{1/M}\big)^{-\hat{k}} \approx (19.5\,M)^{\hat{k}}
\] with probability 0.95, so we truncate at
\(\tau_{\mathrm{eff}} = \min(\tau, \tau_M)\). Truncated at \(\tau_M\),
the tail model describes the runs in which no tail value is above
\(\tau_M\), which are 95\% of the runs under the fitted distribution. A
Pareto distribution truncated in this way has a finite mean and variance
\autocite[App. A]{Vehtari+etal:PSIS:2024}. We combine them with the
empirical mean and variance of the other \(S - M\) values by the law of
total variance, which also accounts for the difference between the two
means, and use the resulting variance
\(\mathrm{Var}_{\mathrm{trunc}}(p(y_i \mid \theta_s))\) in place of the
sample variance and the mean of the combined model,
\(\hat{\mu}_i = (1 - M/S)\,\mu_{\mathrm{bulk}} + (M/S)\,\mu_{\mathrm{tail}}\),
in place of \(\hat{p}_i\), where \(\mu_{\mathrm{bulk}}\) is the mean of
the other \(S - M\) values and \(\mu_{\mathrm{tail}}\) the mean of the
truncated tail, \[
\mathrm{mcse}_{\mathrm{truncGPD},i}
  = \frac{\sqrt{\mathrm{Var}_{\mathrm{trunc}}\big(p(y_i \mid \theta_s)\big)}}{\sqrt{\mathrm{ESS}(a_i)}\;\hat{\mu}_i},
\] where \(\mathrm{ESS}(a_i)\) is the ESS of the values
\(p(y_i \mid \theta_s)\) as in (\ref{eq-delta}), limited to at most
\(S\). We call this \emph{truncated generalized Pareto MCSE} (truncGPD
MCSE). The notebook (Code availability, section \emph{Brute-force})
gives the truncated moments in full.

The largest plausible value \(p_{\max}\) is the smaller of two values: a
value from the data model, and the largest value over the full-data
draws. For discrete data models the value from the data model is a
\emph{physical bound}: 1 for Bernoulli, beta-binomial and
ordered-logistic data, and for Poisson and negative-binomial data the
Poisson probability of \(y_i\) when the mean equals \(y_i\), which is
the largest probability these models can give to \(y_i\). Normal and
Student-\(t\) data models have no physical bound, because their
densities grow without limit when the scale goes to zero. The posterior
usually puts little mass there, so for these models we use a \emph{soft
bound}: the largest value of the density when its scale equals the 5\%
posterior quantile of \(\sigma\). This is not a strict bound, but the
tail model needs only a plausible truncation point, and the results were
not sensitive to the choice of the quantile (notebook, Code
availability, appendix \emph{Largest plausible value \(p_{\max}\):
definitions and comparison}). In LOGO, both are products over the
observations of the group. The second value is the largest value of
\(p(y_i \mid \theta)\) for the same fold \(i\) over the draws
\(\theta^{\mathrm{f}}_s\) from the full-data posterior,
\(\max_s p(y_i \mid \theta^{\mathrm{f}}_s)\). It is never above the
physical bound and usually below the soft bound, so it sets \(p_{\max}\)
when the value from the data model lies far above any value the
posterior supports. Appendix \ref{sec-pmax} provides more details and
compares the two choices.

On the probability scale, the mean is unbiased,
\(\mathrm{E}[\hat p_i] = p(y_i \mid y_{-i})\); its logarithm, the fold
estimate, is not by (\ref{eq-jensen}), \[
\mathrm{E}\big[\widehat{\mathrm{elpd}}_{\mathrm{bf},i}\big] = \mathrm{E}\big[\log \hat p_i\big] \le \log p(y_i \mid y_{-i}) = \mathrm{elpd}_i ,
\] so the bias is downward, and the bias predictor follows from
(\ref{eq-biaspred}) with the truncGPD MCSE. At \(\hat{k} > 1\), both
\(\tau_M\) and the full-data maximum grow with \(S\), and with them the
bias predictor, although the bias itself shrinks. The predictor then
holds only for the \(S\) at which it is computed. Brute-force needs no
additional reliability check, because the truncGPD MCSE already accounts
for a heavy tail.

\par\noindent\colorbox{black!6}{\begin{minipage}{\dimexpr\linewidth-2\fboxsep\relax}

\textbf{Self-diagnostic of brute-force.} \[
\mathrm{mcse}_{\mathrm{bf},i} = \mathrm{mcse}_{\mathrm{truncGPD},i},
\qquad
\mathrm{mcse}_{\mathrm{total},\mathrm{bf}}
  = \sqrt{\textstyle\sum_i \mathrm{mcse}_{\mathrm{bf},i}^2},
\qquad
\mathrm{bias}_{\mathrm{pred},\mathrm{bf},i}
  = -\tfrac{1}{2}\,\mathrm{mcse}_{\mathrm{truncGPD},i}^2 .
\] \emph{Reliability checks:} none.\\
\emph{Trust rule:} \(\mathrm{rmse}_{\mathrm{pred}} < 1\,\mathrm{nat}\).

\end{minipage}}\par

\subsection{PSIS}\label{sec-diag-psis}

For PSIS, the fold estimate is \[
\widehat{\mathrm{elpd}}_{\mathrm{psis},i} = -\log \bar{w}_i,
\qquad
\bar{w}_i = \frac{1}{S}\sum_s w_{i,s},
\qquad
w_{i,s} = \frac{1}{p(y_i \mid \theta_s)},
\] with \(S\) draws from the full-data posterior; that is,
\(a_{i,s} = w_{i,s}\), \(\bar{a}_i = \bar{w}_i\) and
\(\hat{p}_i = 1/\bar{w}_i\) (Section~\ref{sec-background-psis}). In
practice the largest weights are Pareto-smoothed. All fold estimates use
the same draws, so the total MCSE is the IJ total of (\ref{eq-ij}). Let
\(\tilde{w}_{i,s}\) denote the normalized Pareto-smoothed weights of
fold \(i\), so that
\(\hat{p}_i = \sum_s \tilde{w}_{i,s}\, p(y_i \mid \theta_s)\). The
influence of draw \(s\) on fold \(i\) is then \[
\psi_{i,s} = \tilde{w}_{i,s}
  \left( \frac{p(y_i \mid \theta_s)}{\hat{p}_i} - 1 \right),
\qquad \sum_s \psi_{i,s} = 0.
\] A draw under which \(y_i\) is more likely than on average increases
the fold term, and a draw under which it is less likely decreases it.
The influences treat the smoothed weights as fixed, although the
smoothing makes each weight depend on the others. The IJ total is
therefore an approximation, and Appendix \ref{sec-ij-rss} compares it
with the realized SD. The fold MCSE is the plug-in MCSE of
\textcite{Vehtari+etal:PSIS:2024}, computed from the smoothed weights
and taken to the log scale by the delta method (\ref{eq-delta}). The
relative ESS of the influences \(\psi_{i,s}\), \(\mathrm{relESS}_i\),
accounts for the autocorrelation of the draws. The total MCSE follows
from (\ref{eq-ij}) with these influences.

On its own scale, the mean of the raw weights is unbiased,
\(\mathrm{E}[\bar{w}_i] = 1/p(y_i \mid y_{-i})\); the fold estimate is
minus its logarithm, so by (\ref{eq-jensen}), \[
\mathrm{E}\big[\widehat{\mathrm{elpd}}_{\mathrm{psis},i}\big]
  = \mathrm{E}\big[-\log \bar{w}_i\big]
  \geq -\log \frac{1}{p(y_i \mid y_{-i})} = \mathrm{elpd}_i ,
\] and the bias is upward. For the bias predictor we apply the truncated
tail model of brute-force to the raw weights, before Pareto smoothing.
For the weights we have no largest plausible value, so only \(\tau_M\)
applies, and the truncGPD MCSE of the raw weights uses
\(S\,\mathrm{relESS}_i\) in place of \(\mathrm{ESS}(a_i)\). This
truncGPD MCSE is a fold quantity and is used only for the bias
predictor: the bias of the total is the sum of the fold biases, so the
covariances between the folds have no effect on it. The reported MCSE is
the IJ total, which includes the covariances. The reliability check
warns when the Pareto-\(\hat{k}\) of the PSIS weights of a fold,
\(\hat{k}_{\mathrm{psis}}\), is above 0.7
\autocite{Vehtari+etal:PSIS:2024}.

\par\noindent\colorbox{black!6}{\begin{minipage}{\dimexpr\linewidth-2\fboxsep\relax}

\textbf{Self-diagnostic of PSIS.} \[
\mathrm{mcse}_{\mathrm{psis},i}^2
  = \frac{\sum_s \psi_{i,s}^2}{\mathrm{relESS}_i},
\] \[
\mathrm{mcse}_{\mathrm{total},\mathrm{psis}}^2
  = \frac{\sum_s \big(\sum_i \psi_{i,s}\big)^2}{\mathrm{relESS}(\psi)},
\qquad
\mathrm{bias}_{\mathrm{pred},\mathrm{psis},i}
  = +\tfrac{1}{2}\,\mathrm{mcse}_{\mathrm{truncGPD},i}^2 .
\] \emph{Reliability checks:} \(\hat{k}_{\mathrm{psis}} \leq 0.7\) for
every fold.\\
\emph{Trust rule:} \(\mathrm{rmse}_{\mathrm{pred}} < 1\,\mathrm{nat}\)
and every fold passes the reliability check.

\end{minipage}}\par

\subsection{Moment matching}\label{sec-framework-mm}

Moment matching (MM, Section~\ref{sec-background-mm}) can be added to
brute-force and to PSIS, in two different roles.

For brute-force, the draws already come from the leave-fold-out
posterior, so there is no mismatch between proposal and target to
correct. Instead, MM moves the draws toward the region where
\(p(y_i \mid \theta)\) is large, which reduces the error of the mean
\(\hat{p}_i\) when \(y_i\) lies far in the tail
\autocite{Paananen+etal:2021:implicit}. To avoid making an accurate
estimate worse, we apply it only to folds where the self-diagnostic of
the fold warns: when
\(|\mathrm{bias}_{\mathrm{pred},\mathrm{bf},i}| > 1\) nat, or when the
Pareto-\(\hat{k}\) of the values \(p(y_i \mid \theta_s)\),
\(\hat{k}_{\mathrm{bf}}\), is above 1, which shows that the truncGPD
MCSE understates the bias. On a fold with an extreme tail, however, the
shift can move the draws into a region where the leave-fold-out
posterior has little mass. A matched fold therefore keeps its estimate
only if the Pareto-\(\hat{k}\) of the matched terms,
\(\hat{k}_{\mathrm{mm}}\), is at most 0.7, and otherwise returns to the
plain brute-force estimate.

For PSIS, MM transforms the full-data draws of each fold with
\(\hat{k}_{\mathrm{psis}} > 0.7\) and repeats the importance sampling
(Section~\ref{sec-background-mm}). Based on the experimental results, we
use it only as a check, not as an estimator. The Pareto-\(\hat{k}\) of
the new weights, \(\hat{k}_w\), shows whether the leave-fold-out
posterior can be reached from the full-data draws, and a fold with
\(\hat{k}_w \leq 0.7\) loses the warning of plain PSIS. This rule is
empirical. It does not show that the bias predictor of plain PSIS is
accurate on that fold. We report the plain PSIS estimate, for which the
self-diagnostic above is defined. When many folds have
\(\hat{k}_{\mathrm{psis}} > 0.7\), MM cannot help, so we use it as a
check only when fewer than 8\% of the folds do.

\par\noindent\colorbox{black!6}{\begin{minipage}{\dimexpr\linewidth-2\fboxsep\relax}

\textbf{Self-diagnostic with MM.}\\
\emph{Brute-force + MM:} as brute-force, with the truncGPD MCSE of the
matched terms on matched folds; a matched fold is kept only if
\(\hat{k}_{\mathrm{mm}} \leq 0.7\).\\
\emph{PSIS + MM:} estimate, MCSE and bias predictor of plain PSIS; used
as a check only when fewer than 8\% of the folds have
\(\hat{k}_{\mathrm{psis}} > 0.7\).\\
\emph{Trust rule of PSIS + MM:}
\(\mathrm{rmse}_{\mathrm{pred}} < 1\,\mathrm{nat}\), and every fold has
\(\hat{k}_{\mathrm{psis}} \leq 0.7\) or, after matching,
\(\hat{k}_w \leq 0.7\).

\end{minipage}}\par

\subsection{MixIS}\label{sec-diag-mixis}

MixIS samples once from a mixture of all leave-fold-out posteriors, and
its fold estimate has two terms, \[
\widehat{\mathrm{elpd}}_{\mathrm{mixis},i} = \log \bar{u} - \log \bar{v}_i,
\qquad
\bar{u} = \frac{1}{S}\sum_s u_s,
\qquad
\bar{v}_i = \frac{1}{S}\sum_s v_{i,s},
\] with \(S\) draws from the mixture,
\(u_s = \big(\sum_{i'} 1/p(y_{i'} \mid \theta_s)\big)^{-1}\) and
\(v_{i,s} = u_s/p(y_i \mid \theta_s)\). The two means are of the form
\(\bar{a}_i\) with \(a_{i,s} = u_s\) and with \(a_{i,s} = v_{i,s}\), and
\(\hat{p}_i = \bar{u}/\bar{v}_i\). The first term, the \emph{common
term} \(u\), is the same for every fold; the second, the \emph{\(v\)
term}, has bounded weights \(v_{i,s} \in [0, 1]\).

The common term makes the fold errors strongly correlated, so the total
MCSE needs the IJ approach (\ref{eq-ij}). Each draw has the influence
\(\psi_s = \frac{1}{S}\big[n\,(u_s - \bar{u})/\bar{u} - \sum_i (v_{i,s} - \bar{v}_i)/\bar{v}_i\big]\)
on the total, where the first part counts the common term once per fold.
The two means of a fold use the same draws, so the fold MCSE combines
both terms at each draw, and the total MCSE follows from (\ref{eq-ij})
with this influence.

The bias, in contrast, adds up term by term. On their own scales the two
means are unbiased, and
\(\log \mathrm{E}[\bar{u}] - \log \mathrm{E}[\bar{v}_i] = \mathrm{elpd}_i\).
By (\ref{eq-jensen}), the common term \(\log \bar{u}\) is therefore
biased downward and the \(v\) term \(-\log \bar{v}_i\) upward, and by
(\ref{eq-biaspred}) each contributes its own part, with the MCSEs of the
two means, \[
\mathrm{mcse}_u = \frac{\mathrm{sd}(u_s)}{\sqrt{\mathrm{ESS}(u)}\;\bar{u}},
\qquad
\mathrm{mcse}_{v,i} = \frac{\mathrm{sd}(v_{i,s})}{\sqrt{\mathrm{ESS}(v_i)}\;\bar{v}_i},
\] each with the ESS of its own values. Bias and variance therefore
behave differently: the influence of the common term almost cancels in
the total, so its error is nearly pure bias, which the MCSE cannot show
and which enters the predicted RMSE only through the second-order bias
predictor. The predicted RMSE of MixIS is therefore likely to be too
small where the common term is poor. The checks of MixIS concern the
common term. The \(v\) weights are bounded and need no tail check, but
the common weights \(u_s\) are not, and their Pareto-\(\hat{k}\),
\(\hat{k}(u)\), warns above 0.7 that they are heavy-tailed. The MCMC run
on the mixture is checked separately: the mixture can have several
modes, and an ESS of the common weights, \(\mathrm{ESS}(u)\), below
about 10 with four chains shows that the chains are in different modes
\autocite{Vehtari-Gelman-Simpson-etal:2021}.

\par\noindent\colorbox{black!6}{\begin{minipage}{\dimexpr\linewidth-2\fboxsep\relax}

\textbf{Self-diagnostic of MixIS.} \[
\mathrm{mcse}_{\mathrm{mixis},i} =
  \frac{\mathrm{sd}(d_{i,s})}{\sqrt{\mathrm{ESS}(d_i)}},
\qquad
d_{i,s} = \frac{u_s}{\bar{u}} - \frac{v_{i,s}}{\bar{v}_i},
\] \[
\mathrm{mcse}_{\mathrm{total},\mathrm{mixis}}^2
  = \frac{\sum_s \psi_s^2}{\mathrm{relESS}(\psi)},
\qquad
\mathrm{bias}_{\mathrm{pred},\mathrm{mixis},i}
  = -\tfrac{1}{2}\,\mathrm{mcse}_u^2 + \tfrac{1}{2}\,\mathrm{mcse}_{v,i}^2 .
\] \emph{Convergence check (before the trust rule):} MCMC convergence of
the mixture run, including \(\mathrm{ESS}(u)\).\\
\emph{Reliability checks:} \(\hat{k}(u) \leq 0.7\) (warning only).\\
\emph{Trust rule:} \(\mathrm{rmse}_{\mathrm{pred}} < 1\,\mathrm{nat}\).

\end{minipage}}\par

\section{New CV variants}\label{sec-methods-new}

A self-diagnostic tells whether an estimate can be trusted, but it
cannot make the estimate more accurate. This section proposes two new
variants of CV methods. The CV bridge uses all draws of brute-force, and
we propose it as the new reference for evaluating the other methods.
Balanced MixIS makes MixIS more accurate when the folds differ strongly.
For each, we give the estimator and then its self-diagnostic in the
scheme of Section~\ref{sec-framework-diag}.

\subsection{CV bridge}\label{sec-reference-bridge}

Brute-force uses the draws from the leave-fold-out posterior and ignores
the draws from the full-data posterior, although both are available.
Section~\ref{sec-background-bridge} described how bridge sampling with
normal proposals can estimate the normalizing constants of the full-data
posterior and the leave-fold-out posterior. The CV bridge method
estimates \(p(y_i \mid y_{-i})\) directly as the ratio of the
normalizing constants of
\(q_{\mathrm{full}}(\theta) = p(y \mid \theta)\,p(\theta)\) and
\(q_{\mathrm{loo}}(\theta) = p(y_{-i} \mid \theta)\,p(\theta)\). With
\(S\) full-data draws \(\theta^{\mathrm{f}}_s\), \(S\) leave-fold-out
draws \(\theta^{\mathrm{l}}_s\) and the optimal bridge function of
\textcite{Meng+Wong:1996}, the fold estimate is
\begin{equation}\phantomsection\label{eq-bridge}{
\begin{gathered}
\widehat{\mathrm{elpd}}_{\mathrm{bridge},i} = \log \hat{\lambda}_i,
\qquad
\hat{\lambda}_i = \frac{\bar{N}_i}{\bar{D}_i},
\\
N_{i,s} = \frac{p(y_i \mid \theta^{\mathrm{l}}_s)}
  {\omega_{\mathrm{f}}\, p(y_i \mid \theta^{\mathrm{l}}_s)
   + \omega_{\mathrm{l}}\, \hat{\lambda}_i},
\qquad
D_{i,s} = \frac{1}
  {\omega_{\mathrm{f}}\, p(y_i \mid \theta^{\mathrm{f}}_s)
   + \omega_{\mathrm{l}}\, \hat{\lambda}_i},
\end{gathered}
}\end{equation} where \(\bar{N}_i\) and \(\bar{D}_i\) are the means of
\(N_{i,s}\) and \(D_{i,s}\) over the draws, and
\(\hat{p}_i = \hat{\lambda}_i\). Because \(\hat{\lambda}_i\) appears on
both sides, we find it by fixed-point iteration: we start from the
median of \(\log p(y_i \mid \theta^{\mathrm{l}}_s)\) as
\(\log \hat{\lambda}_i\) and stop when \(\log \hat{\lambda}_i\) changes
by less than \(10^{-7}\) between iterations. The weights are the shares
of the two draw sets in their total ESS,
\(\omega_{\mathrm{f}} = \mathrm{ESS}_{\mathrm{f}}/(\mathrm{ESS}_{\mathrm{f}} + \mathrm{ESS}_{\mathrm{l}})\)
and \(\omega_{\mathrm{l}} = 1 - \omega_{\mathrm{f}}\), where
\(\mathrm{ESS}_{\mathrm{f}}\) and \(\mathrm{ESS}_{\mathrm{l}}\) are the
ESS of \(\log p(y_i \mid \theta_s)\) over the full-data draws and over
the leave-fold-out draws. They are computed once, before the iteration.
For autocorrelated draws we use the ESS in place of the draw count,
which \textcite[Sec. 5]{Meng+Wong:1996} conjecture to be the optimal
choice. Up to the fixed-point iteration, the fold estimate is the
difference of two logarithms of means, as in MixIS, with
\(a_{i,s} = N_{i,s}\) and \(a_{i,s} = D_{i,s}\). CV bridge needs both
densities on the same parameter space. We therefore sample each
leave-fold-out posterior with the likelihood weight of the fold set to
zero, which keeps the parameters \(b_g\) of a left-out group in the
model. We evaluate both densities on the unconstrained parameter space,
so that the shift of the MM-mean below cannot move a draw outside the
support. A shift has a unit Jacobian and needs no correction. The CV
bridge then needs only the log densities at both draw sets and a few
iterations, which costs little compared with the MCMC runs of brute
force.

Unlike the values \(p(y_i \mid \theta_s)\) of brute-force, the terms are
bounded, \(N_{i,s} \leq 1/\omega_{\mathrm{f}}\) and
\(D_{i,s} \leq 1/(\omega_{\mathrm{l}} \hat{\lambda}_i)\), so a few
extreme draws cannot dominate the estimate \autocite{Gronau+etal:2017},
and no tail model is needed \autocite{Micaletto+Vehtari:2025}. The fold
MCSE is the delta method (\ref{eq-delta}) applied to the two means, each
with the ESS of its own draws, which gives the relative variance of
\textcite{Micaletto+Vehtari:2025}. The two draw sets are independent, so
the two means have no covariance. The full-data draws, however, enter
every fold through \(D_{i,s}\), so the fold errors are correlated, and
the total MCSE needs the IJ approach (\ref{eq-ij}). The influences
\(\psi^{\mathrm{f}}_{i,s}\) of the full-data draws are summed over the
folds before squaring, as in (\ref{eq-ij}), whereas the influences
\(\psi^{\mathrm{l}}_{i,s}\) of the leave-fold-out draws belong to one
fold each and add as variances. The notebook (Code availability,
appendix \emph{Reference MCSE: IJ derivation}) gives the influences,
which follow from the fixed point of (\ref{eq-bridge}).

By (\ref{eq-biaspred}), each of the two logarithms would contribute a
Jensen term, as in MixIS, with opposite signs. The absolute value of
their sum is below 0.05 nat on a fold that passes the reliability check,
because it is at most about half the squared fold MCSE. On folds that
pass the reliability checks, the CV bridge showed no detectable
finite-\(S\) bias in our experiments
(Section~\ref{sec-reference-check}). We therefore use no bias predictor.
Bridge sampling requires that the two posteriors overlap
\autocite{Meng+Wong:1996,Gronau+etal:2017}, and on folds with a large
influence they may not. We measure the overlap by the
\emph{cross-density gap}, \begin{equation}\phantomsection\label{eq-gap}{
g_i = \Big|\, \mathrm{median}_s \log p(y_i \mid \theta^{\mathrm{l}}_s)
  - \mathrm{median}_s \log p(y_i \mid \theta^{\mathrm{f}}_s) \,\Big| ,
}\end{equation} the distance between the two draw sets on the scale of
the log density ratio
\(\log q_{\mathrm{full}} - \log q_{\mathrm{loo}} = \log p(y_i \mid \theta)\).
We call a gap above 50 nat \emph{wide} and do not use the plain estimate
there. When the two sets do not overlap at all, the iteration does not
converge within 1000 iterations and we return no estimate. We do not use
Pareto-\(\hat{k}\): the terms of (\ref{eq-bridge}) are bounded, so a
tail diagnostic has little to measure, whereas the gap measures directly
what bridge sampling needs.

On folds with a wide gap we move the two posteriors closer before
bridging, as \textcite{Meng+Wong:1996} suggest. The \emph{MM-mean}
shifts the full-data draws, on the unconstrained scale, to the mean of
the leave-fold-out draws,
\(\tilde{\theta}^{\mathrm{f}}_s = \theta^{\mathrm{f}}_s
  - \bar{\theta}^{\mathrm{f}} + \bar{\theta}^{\mathrm{l}}\), the first
step of moment matching (Section~\ref{sec-background-mm}). On folds that
fail the reliability checks, the CV bridge uses the MM-mean instead of
the plain estimate, \[
\widehat{\mathrm{elpd}}_{\mathrm{bridge},i} =
\begin{cases}
  \log \hat{\lambda}_i
    & \text{if finite, } \mathrm{mcse}_{\mathrm{bridge},i} < 0.3
      \text{ and } g_i \leq 50\,\mathrm{nat}, \\
  \log \tilde{\lambda}_i & \text{otherwise},
\end{cases}
\] where \(\tilde{\lambda}_i\) is the estimate after the shift. The
shifted draws have the unnormalized density
\(\tilde{q}_{\mathrm{full}}(\theta) = q_{\mathrm{full}}(\theta
  - \bar{\theta}^{\mathrm{l}} + \bar{\theta}^{\mathrm{f}})\), which has
the same normalizing constant as \(q_{\mathrm{full}}\). Then
\(\tilde{\lambda}_i\) is (\ref{eq-bridge}) with the ratio
\(\tilde{q}_{\mathrm{full}}(\theta)/q_{\mathrm{loo}}(\theta)\) in place
of \(p(y_i \mid \theta)\), evaluated at \(\theta^{\mathrm{l}}_s\) in
\(N_{i,s}\) and at \(\tilde{\theta}^{\mathrm{f}}_s\) in \(D_{i,s}\). We
keep the plain estimate wherever it passes, because the shift adds Monte
Carlo error that the IJ total MCSE does not account for. Unlike
brute-force + bridge sampling
\autocite{Riha-Jedhoff-Buerkner-Vehtari:2026}, which estimates \(p(y)\)
and \(p(y_{-i})\) separately so that their errors add, the CV bridge
estimates their ratio directly, and much of the error cancels. Bridge
sampling of the two marginal likelihoods, each against its own fitted
proposal, does not need the two posteriors to overlap, so we use it as a
fallback where the CV bridge returns no estimate. Its proposals fail for
posteriors of high dimension, and its estimate of \(p(y)\) is shared by
all folds, so the total MCSE must count its variance for every pair of
folds. The RSS of the fold MCSEs would count it only \(n\) times.

\par\noindent\colorbox{black!6}{\begin{minipage}{\dimexpr\linewidth-2\fboxsep\relax}

\textbf{Self-diagnostic of the CV bridge.} \begin{align*}
\mathrm{mcse}_{\mathrm{bridge},i}^2 &=
  \frac{\mathrm{sd}(N_{i,s})^2}{\mathrm{ESS}(N_i)\,\bar{N}_i^2}
  + \frac{\mathrm{sd}(D_{i,s})^2}{\mathrm{ESS}(D_i)\,\bar{D}_i^2},
\\
\mathrm{mcse}_{\mathrm{total},\mathrm{bridge}}^2
  &= \frac{\sum_s \big(\sum_i \psi^{\mathrm{f}}_{i,s}\big)^2}
         {\mathrm{relESS}(\psi^{\mathrm{f}})}
  + \sum_i \frac{\sum_s \big(\psi^{\mathrm{l}}_{i,s}\big)^2}
         {\mathrm{relESS}(\psi^{\mathrm{l}}_i)} .
\end{align*} \emph{Bias predictor:} none.\\
\emph{Reliability checks (per fold):} finite estimate,
\(\mathrm{mcse}_{\mathrm{bridge},i} < 0.3\) and
\(g_i \leq 50\,\mathrm{nat}\).\\
\emph{Trust rule:} a fold is trusted when its estimate passes all
reliability checks. A fold whose plain estimate fails uses the MM-mean,
which is checked again. A run is trusted when every fold is trusted.

\end{minipage}}\par

\subsection{Balanced MixIS}\label{sec-balanced-mixis}

The mixture of MixIS gives fold \(i\) the \emph{mixture share}
\(\pi_i = \mathrm{E}_{q_{\mathrm{mix}}}[v_{i,s}] \propto 1/p(y_i \mid y_{-i})\),
estimated by \(\frac{1}{S}\sum_s v_{i,s}\). When the folds differ
strongly, a few folds therefore dominate the common term and the error
of the estimate. The components of the mixture can have any positive
weights (Section~\ref{sec-background-mixis}). \emph{Balanced MixIS
(bMixIS)} chooses the weights so that every fold has about the same
share, \(\pi_i \approx 1/n\). It samples from
\begin{equation}\phantomsection\label{eq-balanced}{
q^{(\alpha)}_{\mathrm{mix}}(\theta) \propto p(\theta \mid y)
  \sum_{i'} \frac{\alpha_{i'}}{p(y_{i'} \mid \theta)},
\qquad
\alpha_{i} = \exp\!\big(\widehat{\mathrm{elpd}}_{\mathrm{psis},i}
  - \max_{i'} \widehat{\mathrm{elpd}}_{\mathrm{psis},i'}\big),
}\end{equation} where \(\widehat{\mathrm{elpd}}_{\mathrm{psis},i}\) is
the PSIS estimate, already available from the full-data fit. The weights
become \(u_s = \big(\sum_{i'} \alpha_{i'}/p(y_{i'} \mid
\theta_s)\big)^{-1}\) and
\(v_{i,s} = u_s\,\alpha_i/p(y_i \mid \theta_s)\), so the \(v\) weights
stay in \([0, 1]\), and the estimate remains consistent for any positive
\(\alpha_i\). The fold estimate is \[
\widehat{\mathrm{elpd}}_{\mathrm{bmixis},i}
  = \log \bar{u} - \log \bar{v}_i + \log \alpha_i ,
\] because
\(\mathrm{E}[v_{i,s}]/\mathrm{E}[u_s] = \alpha_i / p(y_i \mid y_{-i})\).
With \(\pi_i \propto \alpha_i/p(y_i \mid y_{-i})\), a fold that is hard
to predict, with a small \(p(y_i \mid y_{-i})\), gets a small
\(\alpha_i\), and the shares become about equal.

The self-diagnostic of bMixIS is that of MixIS with these \(u_s\) and
\(v_{i,s}\). In addition, the importance sampling ESS of the common
weights, \(\mathrm{ESS}_{\mathrm{IS}}(u)\), can be predicted before the
mixture is sampled. With
\(m_\alpha(\theta) = \sum_{i'} \alpha_{i'}/p(y_{i'} \mid \theta)\), \[
\mathrm{ESS}_{\mathrm{IS}}(u) \approx S\,\mathrm{relESS}_{\mathrm{IS}}(\alpha),
\qquad
\mathrm{relESS}_{\mathrm{IS}}(\alpha) =
  \frac{1}{\mathrm{E}_{p(\theta \mid y)}[m_\alpha]\;
           \mathrm{E}_{p(\theta \mid y)}[1/m_\alpha]},
\] computed from the full-data draws alone. The Pareto-\(\hat{k}\) of
\(m_\alpha\) warns when this prediction is itself unreliable. The ratio
\(\mathrm{relESS}_{\mathrm{IS}}(\alpha)/\mathrm{relESS}_{\mathrm{IS}}(1)\)
shows before sampling how much balancing is likely to help. With a ratio
near 1, balancing is unlikely to help. The prediction has no threshold,
because what counts as a small ESS depends on the case, and it is not
part of the trust rule. We balance by default and use the ratio only to
save computation when draws are expensive.

\par\noindent\colorbox{black!6}{\begin{minipage}{\dimexpr\linewidth-2\fboxsep\relax}

\textbf{Self-diagnostic of bMixIS.} MCSE, bias predictor, reliability
checks and trust rule as for MixIS, with \(u_s\) and \(v_{i,s}\) from
(\ref{eq-balanced}).\\
\emph{Optional prediction before sampling:}
\(\mathrm{ESS}_{\mathrm{IS}}(u) \approx S\,\mathrm{relESS}_{\mathrm{IS}}(\alpha)\),
with no threshold. The Pareto-\(\hat{k}\) of \(m_\alpha\) warns when the
prediction is unreliable.

\end{minipage}}\par

\section{Integrating out group-specific
parameters}\label{sec-integration}

Integration is not a way to compute CV but a change to what is computed,
and it can be combined with any of the methods of
Section~\ref{sec-background}. We include integration of group-specific
parameters to increase the variability in the behavior of examples in
our experiments. Appendix \ref{sec-cells} explains how we selected the
experimental cells. Both brute-force and importance sampling handle the
group-specific parameters \(b_g\) by sampling them, together with
\(\phi\), and the CV computation uses \emph{conditional} terms
\(p(y_g \mid b_g, \phi)\). An \emph{integrated} computation uses terms
\[
  p(y_g \mid \phi) = \int p(y_g \mid b_g, \phi)\, p(b_g \mid \phi)\, db_g,
\] where the group-specific parameters \(b_g\) have been integrated out.

In LOGO the parameters \(b_g\) of the left-out group are informed only
by the population distribution, so their leave-fold-out posterior is far
from their full-data posterior, and importance weights that depend on
\(b_g\) tend to have heavy tails. Brute-force has a difficulty of its
own. The refit without group \(g\) contains no data from the group, so
it gives draws of \(\phi\) but no information about \(b_g\). The usual
refit therefore draws \(b_g\) from \(p(b_g \mid \phi)\) for each draw of
\(\phi\) and averages \(p(y_g \mid b_g, \phi)\) over these draws
\autocite{Riha-Jedhoff-Buerkner-Vehtari:2026}. When the group has many
observations, \(p(y_g \mid b_g, \phi)\) is large only in a small region
of \(b_g\), few draws from \(p(b_g \mid \phi)\) fall into this region,
and the average is dominated by these few draws
\autocite{Gronau+etal:2017}. LOO in models with singleton or small
groups has the same computational problem. For LOO with several
observations per group, the integrated per-observation predictive
density is a marginal-likelihood ratio. For observation \(i\) in group
\(g\), per global draw \(\phi_s\), \[
  p(y_i \mid y_{g \setminus i}, \phi_s) =
\frac{\int p(y_g \mid \phi_s, b_g)\, p(b_g \mid \phi_s)\, db_g}{\int p(y_{g \setminus i} \mid \phi_s, b_g)\, p(b_g \mid \phi_s)\, db_g}.
\] In most cases we compute these integrals after MCMC, from the saved
draws, so that integration can be added only when a self-diagnostic
warns.

For normal data models this integral has a closed form, and
\textcite{Vehtari+etal:2016:LOO_for_GLVM} call this \emph{integrated
importance sampling}. For other data models the integral must be
approximated. \textcite{Held-Schrodle-Rue:2010} use \emph{Laplace
approximation}, which replaces the integrand by a normal approximation
around its mode. \textcite{Li-Qiu-Zhang-etal:2016} use simple Monte
Carlo by sampling independently from \(p(b_g \mid \phi)\).
\textcite{Merkle+Furr+Rabe-Hesketh:2019} use \emph{adaptive
Gauss--Hermite quadrature}, a rule that evaluates the integrand at a few
fixed points, placed according to the posterior mean and standard
deviation of the group's parameters. \textcite{Bindoff:2026} integrates
a scalar group effect in closed form for normal data models and
otherwise by quadrature over the population distribution of the effect,
with 64 points scaled by its standard deviation.

\textcite{Riha-Jedhoff-Buerkner-Vehtari:2026} compare simple Monte
Carlo, adaptive Gauss--Hermite quadrature, bridge sampling, and the
Laplace approximation, each combined with PSIS. Every method that uses
an approximate integral inherits the error of that approximation, a
point we return to in Section~\ref{sec-discussion}. The other integrated
methods (integrated brute-force, integrated MixIS) are named in the same
way as integrated PSIS.

The integration rule depends on the data model and the dimension of
\(b_g\). For normal data models with \(b_g\) in the mean, the integral
is analytic. Where the integrand is not log-concave, as can happen with
Student-\(t\) and beta-binomial data models, the Laplace approximation
can fail, and quadrature is feasible for one or two parameters. Where
the integrand is (nearly) log-concave, the Laplace approximation can be
corrected by importance sampling, also in more dimensions. We propose an
efficient approach for that correction. The proposal distribution is a
split-normal widened separately on each side of the mode
\autocite{Geweke:1989} with a widening rule modified to exploit the fact
that the prior is normal. Instead of independent draws we use randomized
quasi-Monte Carlo \autocite{LEcuyer:2018:rqmc,Owen:2026:rqmc}.
Independent randomizations of the point set are used to estimate the
Jensen bias of the logarithm of the inner estimate. Unlike the bias of
the fold estimate (Section~\ref{sec-framework-diag}), we remove this
bias, because the inner integrand is light-tailed, the second-order term
is therefore accurate, and even small bias would accumulate when
averaging over many posterior draws. The notebook (Code availability,
appendices \emph{Integration rules for the group-specific parameters}
and \emph{Split-RQMC IS}) describes the correction and its remaining
bias.

\section{Experiments}\label{sec-experiments}

In this section, we run repeated experiments to check whether the
self-diagnostics of Section~\ref{sec-framework} are calibrated, that is,
whether the error they predict from a single run matches the error that
the CV methods actually make. We compute each method 100 times on up to
37 combinations of models and data sets and compare its estimates with
an accurate reference. We start where the motivation of this paper
starts: brute-force is the usual reference, but it can be biased. To
measure its error at all, we first need a more accurate reference. Q1
therefore checks whether the CV bridge is accurate enough to serve as
the reference, and all later results rely on it. Q2 then asks whether
brute-force is accurate, and whether its self-diagnostic flags the cases
where it is not. Q3 evaluates the fast methods, and Q4 asks whether
balancing makes MixIS more accurate.

\subsection{Design}\label{sec-testbed-cells}

We chose 37 combinations of a model and a data set, \emph{cells} in what
follows, that cover both LOO (19 cells) and LOGO (18 cells). The data
models include normal, Student-\(t\), Poisson, negative-binomial,
beta-binomial, Bernoulli and ordered-logistic models and a Gaussian
process regression, and most of the models are hierarchical. A
hierarchical model can also be computed with the parameters \(b_g\) of
each group integrated out, so for 13 of the hierarchical models we
introduce two cells: In the \emph{base cell}, \(b_g\) is sampled, as in
the original model. In the \emph{integrated cell}, \(b_g\) is integrated
out, and the cell is labeled by its integration rule: analytic (int-an),
quadrature (int-qu), analytic integration of the mean with
one-dimensional quadrature (int-anqu), or the Laplace approximation with
an importance sampling correction (int-srqis). Appendix \ref{sec-cells}
lists all cells with their data models, numbers of folds, posterior
dimensions and references (Table~\ref{tbl-cells}), and the integration
rule of each integrated LOGO cell (Table~\ref{tbl-int-cells}).

The design is not fully crossed: not every method is computed on every
cell. Integrated brute-force is computed on all 13 integrated cells.
PSIS + MM is computed on the LOO base cells and on 9 of the 13
integrated cells. Brute-force + MM is computed on every base cell except
verbagg\_grp, and we report it on the cells where the brute-force trust
rule rejects at least one run, because we apply moment matching to
brute-force only where its self-diagnostic warns
(Section~\ref{sec-framework-mm}). The notebook (Code availability,
appendix \emph{Per-method per-cell result tables}) lists the cells of
each method.

As reference of each cell we use the CV bridge of
Section~\ref{sec-methods-new}, computed with \(S = 16000\) draws on each
side and averaged over 100 runs. As its trust rule prescribes, a fold
uses the plain estimate when it passes the reliability checks and the
MM-mean otherwise. Six cells deviate from this rule, and Appendix
\ref{sec-cells} lists them with the reasons.

Each cell is computed \(R = 100\) times with the same data and a new
random seed for MCMC, so that the variation over runs is Monte Carlo
error only. Every method uses \(S = 4000\) posterior draws from 4 chains
of the No-U-Turn sampler, the default of Stan
\autocite{Hoffman+Gelman:2014:nuts,StanTeam:2026}, initialized with
Pathfinder \autocite{Zhang+etal:2022:pathfinder}, which reduces the
chance that a chain starts in a minor mode. For run \(r\) of a method,
the error is
\(\mathrm{err}_r = \widehat{\mathrm{elpd}}_m^{(r)} - \mathrm{elpd}_{\mathrm{ref}}\),
where \(\mathrm{elpd}_{\mathrm{ref}}\) is the mean of the 100 reference
runs. Table~\ref{tbl-measures} lists what we compare: each quantity that
the self-diagnostic predicts from one run, and its realized counterpart
over the 100 runs. A self-diagnostic is calibrated when the two agree.

\begin{longtable}[]{@{}
  >{\raggedright\arraybackslash}p{(\linewidth - 4\tabcolsep) * \real{0.2027}}
  >{\raggedright\arraybackslash}p{(\linewidth - 4\tabcolsep) * \real{0.3378}}
  >{\raggedright\arraybackslash}p{(\linewidth - 4\tabcolsep) * \real{0.4595}}@{}}
\caption{What we compare for every method and
cell.}\label{tbl-measures}\tabularnewline
\toprule\noalign{}
\begin{minipage}[b]{\linewidth}\raggedright
\end{minipage} & \begin{minipage}[b]{\linewidth}\raggedright
Predicted in each run
\end{minipage} & \begin{minipage}[b]{\linewidth}\raggedright
Realized over the 100 runs
\end{minipage} \\
\midrule\noalign{}
\endfirsthead
\toprule\noalign{}
\begin{minipage}[b]{\linewidth}\raggedright
\end{minipage} & \begin{minipage}[b]{\linewidth}\raggedright
Predicted in each run
\end{minipage} & \begin{minipage}[b]{\linewidth}\raggedright
Realized over the 100 runs
\end{minipage} \\
\midrule\noalign{}
\endhead
\bottomrule\noalign{}
\endlastfoot
Random error & \(\mathrm{mcse}_{\mathrm{total}}\) & realized SD:
\(\mathrm{sd}_r(\mathrm{err}_r)\) \\
Bias & \(\mathrm{bias}_{\mathrm{pred}}\) & realized bias:
\(\frac{1}{R}\sum_r \mathrm{err}_r\) \\
Total error & \(\mathrm{rmse}_{\mathrm{pred}}\) & realized RMSE:
\(\big(\frac{1}{R}\sum_r \mathrm{err}_r^2\big)^{1/2}\) \\
Interval & asymmetric and symmetric 95\% interval & coverage: share of
runs with \(\mathrm{err}_r\) inside \\
Decision & trust rule & trusted \%: share of runs that are trusted \\
\end{longtable}

We fit the models with CmdStan through cmdstanr
\autocite{Gabry-Cesnovar-Johnson-etal:2024} and brms
\autocite{Buerkner:2017:brms}, compute PSIS and MM for PSIS with the loo
package \autocite{Vehtari+etal:2026:loo}, Pareto-\(\hat{k}\) with the
posterior package \autocite{Burkner-Gabry-Kay-etal:2026}, and MM for
brute-force with the iwmm package of
\textcite{Paananen+etal:2021:implicit}. All methods are implemented in
the R package loobench (see \url{https://github.com/avehtari/loobench}).

\subsection{Q1: Is the CV bridge an accurate
reference?}\label{sec-reference-check}

\subsubsection{Is the CV bridge accurate enough to be the
reference?}\label{is-the-cv-bridge-accurate-enough-to-be-the-reference}

For the CV bridge to be a reasonable reference, both its bias and its
random error must be small. By (\ref{eq-delta}) and (\ref{eq-biaspred}),
the finite-\(S\) bias of a logarithm of a Monte Carlo mean is
proportional to \(1/S\) when the values have finite variance, as the
bounded terms of the CV bridge do. A relevant bias would therefore
change the mean between \(S = 4000\) and the \(S = 16000\) of the
references. On all 24 references (one per base cell, as an integrated
cell shares the reference of its base cell), the change in the mean is
at most 0.13 nat (0.31 nat on verbagg\_grp). The bias at \(S = 16000\)
is then one third of this change, that is, below 0.05 nat (about 0.1 nat
on verbagg\_grp). Both values are small compared with the 1-nat
threshold of the trust rule. As this check cannot see a bias that does
not shrink with \(S\), we also compare CV bridge with integrated PSIS.
Both methods share no computation, so they are unlikely to have the same
bias. Integrated PSIS agrees with the references to within 0.38 nat on
every integrated cell except roaches, whose reference it is, and
pupil\_grp, where it fails (for details, see
Section~\ref{sec-fast-psis}). A bias that does not shrink with \(S\) is
therefore also small against the 1-nat threshold on these cells. On
roaches, the CV bridge between the marginal posteriors, computed with
\(S = 4000\), agrees with the reference to 0.009 nat, which supports
integrated PSIS as the reference there. The standard error of each
reference, the SD of its 100 runs divided by 10, is below 0.06 nat. The
error \(\mathrm{err}_r\) of a run is therefore known to about 0.1 nat.
This is precise enough to tell whether it is above or below the 1-nat
threshold of the trust rule.

Both the bias and the random error of the references are well below the
1-nat threshold of the trust rule. The CV bridge is therefore accurate
enough to be the reference on 21 of the 24 base cells. On roaches, we
use integrated PSIS. On pupil\_sigma\_grp, we use the CV bridge on the
integrated model, and on pupil\_grp, bridge sampling of the two marginal
likelihoods of the integrated model (Table~\ref{tbl-cells}).

\subsubsection{Does the CV bridge report its own error
correctly?}\label{does-the-cv-bridge-report-its-own-error-correctly}

\begin{figure}[t]

\centering{

\includegraphics[width=1\linewidth,height=\textheight,keepaspectratio]{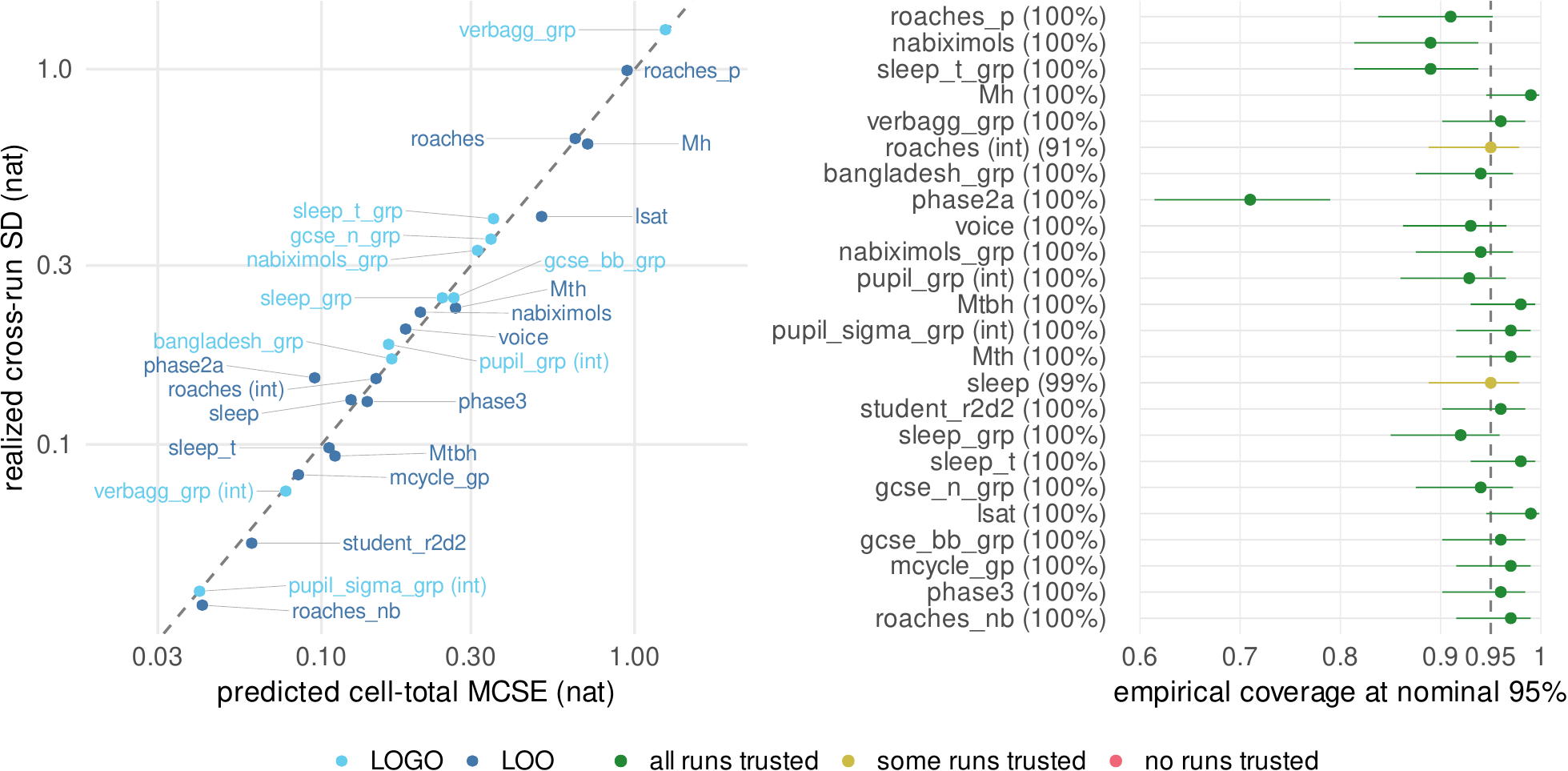}

}

\caption{\label{fig-ref-calib}The self-diagnostic of the reference
methods is calibrated on most cells (\(S = 4000\)). Left: IJ total MCSE
against the realized SD over 100 runs, one point per cell; dashed line:
identity. Right: share of runs in which the symmetric 95\% interval
contains the error, with Wilson intervals; dashed line: 0.95; trusted \%
after each cell name.}

\end{figure}

A user computes the CV bridge once and must rely on its MCSE. To check
the MCSE, we computed each reference method also with \(S = 4000\) in
each of the 100 runs, from the draws of brute-force. We compared these
runs with the mean of the same method at \(S = 16000\). In
Figure~\ref{fig-ref-calib}, the left panel plots the IJ total MCSE
against the realized SD. A point above the dashed diagonal means that
the MCSE is too small. The right panel shows the share of runs in which
the symmetric 95\% interval contains the error and the color gives the
trusted \%. Most points lie close to the diagonal, and most intervals
contain the error in close to 95\% of the runs. The exception is
phase2a, the only cell where every fold uses the MM-mean. Its point lies
above the diagonal, and its interval contains the error in only 71\% of
the runs. The IJ misses the Monte Carlo error of the shift
(Section~\ref{sec-reference-bridge}), so the MCSE of the MM-mean is not
reliable. The reference itself, which averages 100 runs, is not
affected. On roaches\_p, two or three folds fail the checks of the plain
estimate in every run, but the MM-mean passes on these folds, so every
run is trusted. The MCSE of the CV bridge is therefore calibrated on
every cell except phase2a. On phase2a, the MCSE is too small, but the
reference is not affected.

\subsubsection{Does its self-diagnostic flag when the CV bridge cannot
be
trusted?}\label{does-its-self-diagnostic-flag-when-the-cv-bridge-cannot-be-trusted}

As discussed above, the CV bridge fails on two cells: pupil\_grp and
pupil\_sigma\_grp. On both cells, the two posteriors of the conditional
model overlap on only 1 or 2 of the 20 folds. The checks reject the
other folds, so we compute the CV bridge on the integrated model. On
pupil\_sigma\_grp, this is the reference. On the integrated model of
pupil\_grp, leaving out subject 20, with 120 observations, moves the
posterior so far that the gap on its fold is about 120 nat, against
about 1.7 on the other folds. The gap check (\(g_i \leq 50\) nat)
rejects exactly this fold, in every run. The reference of this cell
therefore comes from bridge sampling of the two marginal likelihoods,
which does not need the posteriors to overlap.

On phase2a, the gap of one fold is about 210 nat. The gap check rejects
this fold in every run, and its MM-mean passes the checks. With the
trust rule at \(S = 4000\), the realized RMSE is 0.66 nat, below the
1-nat threshold. The median IJ total MCSE is 0.63 nat, and the realized
SD is 0.66 nat.

\par\noindent\colorbox{black!6}{\begin{minipage}{\dimexpr\linewidth-2\fboxsep\relax}

\textbf{Answer to Q1.} Yes. The errors of the CV bridge references are
small compared with 1 nat. Its MCSE is calibrated, except on folds that
use the MM-mean, and its gap check finds the folds where the CV bridge
fails.

\end{minipage}}\par

\subsection{Q2: Is brute-force a good reference?}\label{sec-bruteforce}

\subsubsection{How accurate is
brute-force?}\label{how-accurate-is-brute-force}

Brute-force is accurate when the data model fits the data and no fold
has an extreme influence. On 18 of the 19 base cells of this kind, its
realized bias is at most 0.36 nat and its realized RMSE is below 1 nat.
For roaches, the bias is small (\(-0.65\) nat), but the random error is
large (SD 1.1 nat). However, brute-force fails in three scenarios:
\emph{Too-narrow data model:} a few observations lie far in the tail of
their predictive distribution. Their values \(p(y_i \mid \theta_s)\)
then have a heavy upper tail, and the fold estimate is biased downward
(Section~\ref{sec-framework-diag}). This is the case for roaches\_p
(\(-17\) nat) and phase2a (\(-78\) nat). \emph{Many folds:} the bias of
the elpd is the sum of the fold biases, so small fold biases add up.
This is the case for verbagg\_grp, with 316 folds and an error of \(-5\)
nat, although no single fold has a large error. \emph{Influential
groups:} in LOGO, brute-force draws the parameters of the left-out group
from their population distribution. When a group has many observations,
few draws fall where its data are likely
(Section~\ref{sec-integration}). This is the case for the two pupil
cells, with up to 120 observations per group and errors of about \(-48\)
nat.

\subsubsection{Does its self-diagnostic flag when brute-force is
wrong?}\label{does-its-self-diagnostic-flag-when-brute-force-is-wrong}

\begin{figure}[t]

\centering{

\includegraphics[width=1\linewidth,height=\textheight,keepaspectratio]{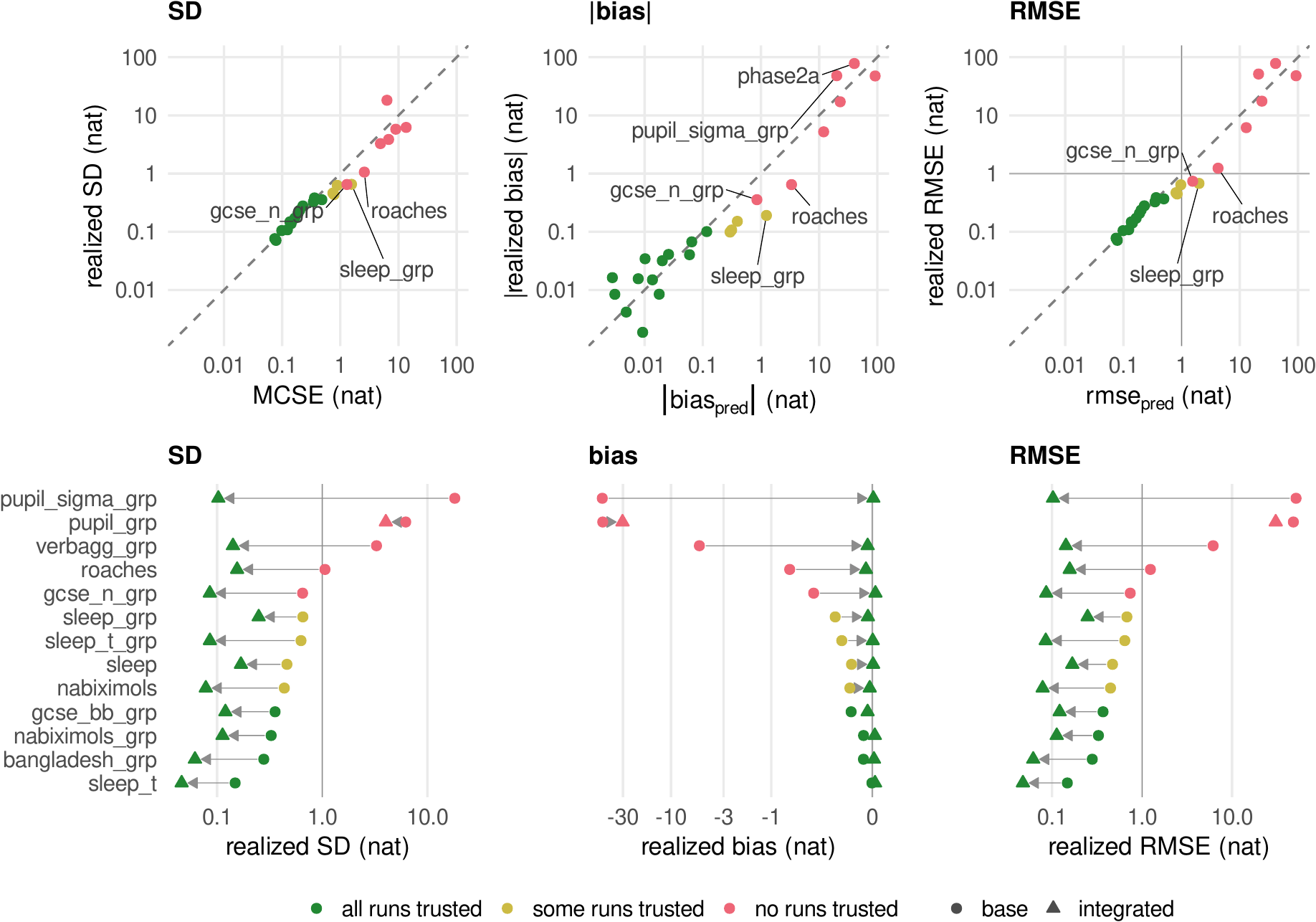}

}

\caption{\label{fig-bf-calib}The brute-force self-diagnostic is
calibrated where brute-force works and rejects every cell with a large
error, and integration removes most of the error on 12 of 13 cells.
\(S = 4000\), colored by trusted \%. Top: predicted against realized SD,
absolute bias and RMSE, on one log scale, one point per base cell;
dashed line: identity. Bottom: realized SD, bias and RMSE over the 100
runs for each base cell (circle) and its integrated cell (triangle); the
arrow points from the base cell to the integrated cell; thin solid line:
1 nat (SD, RMSE) and 0 (bias).}

\end{figure}

The top row of Figure~\ref{fig-bf-calib} compares, for each base cell,
the predicted MCSE, bias and RMSE with their realized values over the
100 runs, on one log scale. In all three panels, a point above the
diagonal means that the self-diagnostic underestimates the error, and
the color gives the trusted \%. The bottom row shows the integrated
cells. Every cell with a large error is red and lies at a realized RMSE
above 1 nat in the top right panel, so the trust rule rejects all of its
runs. On 16 of the 18 base cells where brute-force is accurate,
82--100\% of runs are trusted. The exceptions are gcse\_n\_grp and
sleep\_grp (realized RMSE 0.74 and 0.68 nat), where the trust rule
rejects 100\% and 98\% of the runs. On roaches, rejecting every run is
right: its realized RMSE is 1.2 nat, because of the large random error.

\subsubsection{Does the self-diagnostic give the right sign and
magnitude of the
error?}\label{does-the-self-diagnostic-give-the-right-sign-and-magnitude-of-the-error}

\begin{figure}[t]

\centering{

\includegraphics[width=1\linewidth,height=\textheight,keepaspectratio]{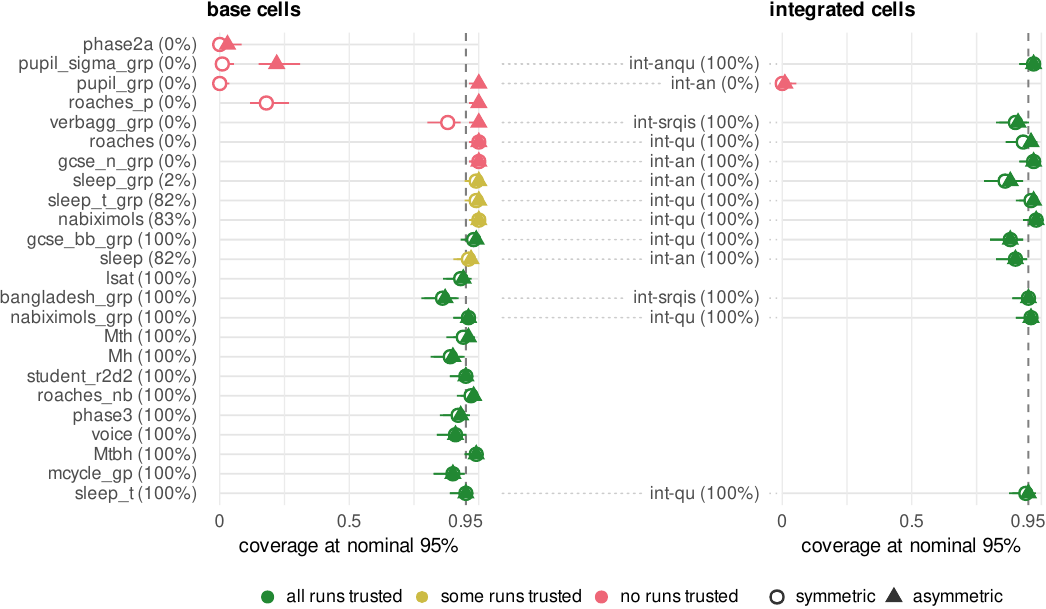}

}

\caption{\label{fig-bf-cover}The asymmetric interval improves coverage
where the bias predictor is roughly right. Share of runs in which the
95\% interval of brute-force contains the error: asymmetric interval
(triangle) and symmetric interval (circle), with Wilson intervals;
dashed line: 0.95; trusted \% after each cell name.}

\end{figure}

Where brute-force is accurate, its truncGPD MCSE is close to the
realized SD on most cells (top left panel of Figure~\ref{fig-bf-calib}).
On five cells it is 1.4 to 2.4 times larger, which is the safe
direction. The sign of the bias predictor is right: wherever brute-force
has a relevant bias, it is downward, as predicted.
Figure~\ref{fig-bf-cover} shows the share of runs in which the symmetric
interval (circle) and the asymmetric interval (triangle) contain the
error. On most cells the two markers coincide, because the bias
predictor is small. On pupil\_grp and roaches\_p, however, the symmetric
interval contains the error in 0\% and in 18\% of the runs, while the
asymmetric interval, wider on the side of the downward bias, contains it
in every run.

The magnitude of the bias predictor is not always right. In the top
middle panel of Figure~\ref{fig-bf-calib}, a point above the diagonal
shows a bias predictor that is too small. This is the case on phase2a
(\(-40\) against a realized \(-78\) nat), where the full-data posterior
puts almost no mass where the left-out observation is likely, and on
pupil\_sigma\_grp (\(-20\) against \(-48\) nat). More draws do not
remove this bias: on phase2a it is still \(-69\) nat with \(S = 16000\).
On roaches\_p, verbagg\_grp and pupil\_grp, the bias predictor is too
large instead.

\subsubsection{Can brute-force be
repaired?}\label{can-brute-force-be-repaired}

On the hierarchical cells, integration (Section~\ref{sec-integration})
helps. In the bottom row of Figure~\ref{fig-bf-calib}, an arrow that
points left, or towards zero for the bias, shows a smaller error.
Integration reduces the RMSE from 0.74 to 0.09 nat on gcse\_n\_grp and
from 0.68 to 0.25 nat on sleep\_grp, and it removes the whole error on
pupil\_sigma\_grp (\(-48\) to \(0.00\) nat). On pupil\_grp it removes
less than half (\(-48\) to \(-30\) nat), because the heavy tail is now
in \(p(y_g \mid \phi)\) as a function of \(\phi\), and the trust rule
still rejects every run. On the cells with a too-narrow data model,
moment matching (Section~\ref{sec-framework-mm}), applied only to the
folds where the self-diagnostic warns, reduces the error from \(-17.1\)
to \(-0.55\) nat on roaches\_p and from \(-78\) to \(-0.75\) nat on
phase2a (Appendix \ref{sec-bf-mm}). A user who has run brute-force,
however, does not need it. The CV bridge is computed from the same
draws, without new MCMC runs, and it is more accurate or as accurate on
these cells (Section~\ref{sec-reference-check}). We therefore recommend
the CV bridge instead of moment matching.

\par\noindent\colorbox{black!6}{\begin{minipage}{\dimexpr\linewidth-2\fboxsep\relax}

\textbf{Answer to Q2.} Not on its own. Brute-force can be strongly
biased when the data model is too narrow, when there are many folds, or
when a left-out group is influential. Its self-diagnostic flags these
cases. Integration repairs some of them, and the CV bridge, computed
from the same draws, is more accurate.

\end{minipage}}\par

\subsection{Q3a: How good is PSIS?}\label{sec-fast-psis}

\subsubsection{How accurate is PSIS?}\label{how-accurate-is-psis}

\begin{figure}[t]

\centering{

\includegraphics[width=1\linewidth,height=\textheight,keepaspectratio]{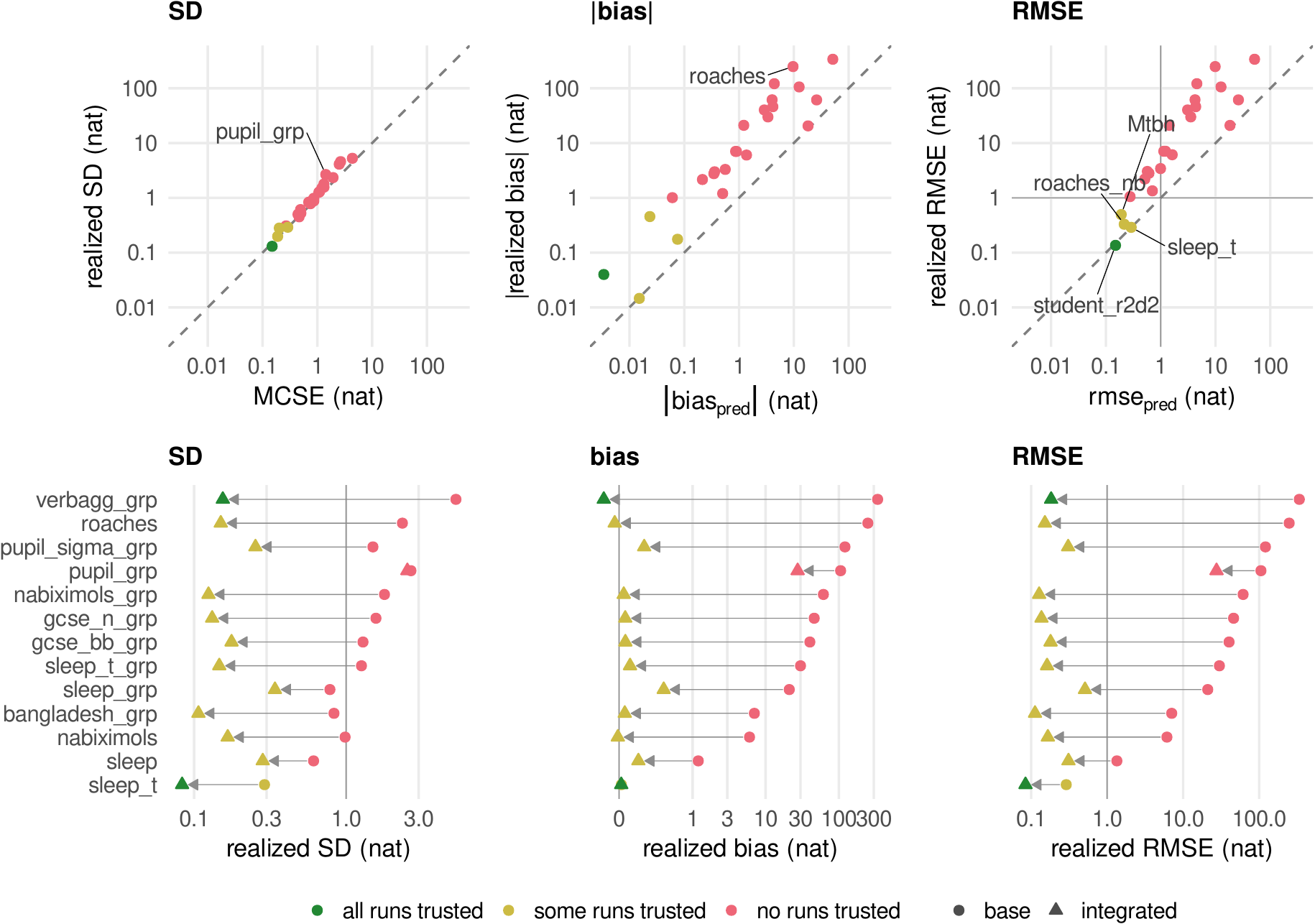}

}

\caption{\label{fig-psis-calib}Above \(\hat{k}_{\mathrm{psis}} = 0.7\)
the PSIS bias predictor is mostly 4 to 17 times too small. PSIS at
\(S = 4000\), colored by trusted \%. Top: predicted against realized SD,
absolute bias and RMSE, on one log scale, one point per base cell;
dashed line: identity. Bottom: realized SD, bias and RMSE over the 100
runs for each base cell (circle) and its integrated cell (triangle); the
arrow points from the base cell to the integrated cell; thin solid line:
1 nat (SD, RMSE) and 0 (bias).}

\end{figure}

Figure~\ref{fig-psis-calib} has the same layout as
Figure~\ref{fig-bf-calib}. PSIS is accurate where the realized bias (top
middle panel) and the realized RMSE (top right panel) are small. Among
the base cells, this is the case on sleep\_t, student\_r2d2, roaches\_nb
and Mtbh, with a realized RMSE of at most 0.5 nat. On sleep\_t and
student\_r2d2, every fold has \(\hat{k}_{\mathrm{psis}} \leq 0.7\) in
almost every run. On the other base cells, many folds have
\(\hat{k}_{\mathrm{psis}} > 0.7\): the full-data posterior is a poor
proposal for their leave-fold-out posterior, and the weights have a
heavy tail (Section~\ref{sec-background-psis}). PSIS is then strongly
biased upward. Based on these results, it may look like PSIS is rarely
accurate, but we intentionally selected many hard cells, and in practice
PSIS is often accurate enough.

\subsubsection{Does its self-diagnostic flag when PSIS is
wrong?}\label{does-its-self-diagnostic-flag-when-psis-is-wrong}

Yes. In Figure~\ref{fig-psis-calib}, every cell with a large error is
red, and the trust rule rejects its runs. On most of them, the predicted
RMSE alone is already above 1 nat. On six cells (Mth, phase3, sleep,
voice, lsat and mcycle\_gp), however, the predicted RMSE is below 1 nat,
although the realized RMSE is 1.1 to 3.4 nat. These red cells lie left
of 1 nat in the top right panel, and only the reliability check
\(\hat{k}_{\mathrm{psis}} \leq 0.7\) rejects their runs. The same check
also rejects some accurate runs: on Mtbh and roaches\_nb (realized RMSE
0.5 and 0.33 nat), only 1\% and 11\% of runs are trusted, and only 5
integrated cells are trusted in 91--100\% of runs.

\subsubsection{Does the self-diagnostic give the right sign and
magnitude of the
error?}\label{does-the-self-diagnostic-give-the-right-sign-and-magnitude-of-the-error-1}

Where PSIS is accurate, the total MCSE lies close to the realized SD
(top left panel). On the cells with many folds with
\(\hat{k}_{\mathrm{psis}} > 0.7\), it is too small by up to a factor of
two, for example 1.4 against 2.7 nat on pupil\_grp; these cells lie
slightly above the diagonal. Counting the covariance between folds
matters, because all folds use the same draws. Without it, the total
MCSE is 0.58--0.95 times the realized SD, and the 95\% interval contains
the error in only 81\% of the runs (Appendix \ref{sec-ij-rss}). The
direction of the bias predictor is right: on every cell with a relevant
bias, both the predicted and the realized bias are positive. Its
magnitude, however, is usually far too small where
\(\hat{k}_{\mathrm{psis}} > 0.7\), mostly by a factor of 4 to 17 and by
a factor of up to about 28 on pupil\_sigma\_grp; only on roaches\_p is
it about right (18 against 21 nat). On roaches, for example, it predicts
about 10 nat against a realized 249 nat; such cells lie far above the
diagonal in the middle panel of Figure~\ref{fig-psis-calib}. This agrees
with \textcite{Vehtari+etal:PSIS:2024}: above 0.7, the bias of PSIS
dominates its variance. More draws do not remove this bias: from
\(S = 4000\) to \(S = 16000\), only roaches\_nb and lsat reduce their
bias at the Monte Carlo rate. On the cells with a large bias, both
intervals rarely contain the error.

\subsubsection{Can PSIS be repaired?}\label{can-psis-be-repaired}

Integration removes most of the error. The bottom row of
Figure~\ref{fig-psis-calib} shows each base cell with its integrated
cell. Every integrated cell except pupil\_grp has a bias of at most 0.38
nat, against \(+7\) to \(+340\) nat for the same LOGO base cells. On
roaches (int), integrated PSIS is its own reference, so its error there
shows only the variation between runs. On pupil\_grp, integration
repairs the weights but not the strong influence of the left-out group,
and its bias of \(+27\) nat is rejected in every run.

\par\noindent\colorbox{black!6}{\begin{minipage}{\dimexpr\linewidth-2\fboxsep\relax}

\textbf{Answer to Q3a PSIS.} PSIS is accurate when every
\(\hat{k}_{\mathrm{psis}}\) is at most 0.7. It can be accurate with a
few higher \(\hat{k}_{\mathrm{psis}}\), but the trust rule then rejects
the run. Above 0.7, its bias predictor is usually far too small, but the
reliability check flags these cases. Its IJ total MCSE is calibrated
where PSIS is accurate. Integration makes most LOGO cells accurate.

\end{minipage}}\par

\subsection{Q3b: How good is PSIS + MM}\label{sec-fast-mm}

\subsubsection{Does moment matching improve the PSIS
estimate?}\label{does-moment-matching-improve-the-psis-estimate}

\begin{figure}[t]

\centering{

\includegraphics[width=1\linewidth,height=\textheight,keepaspectratio]{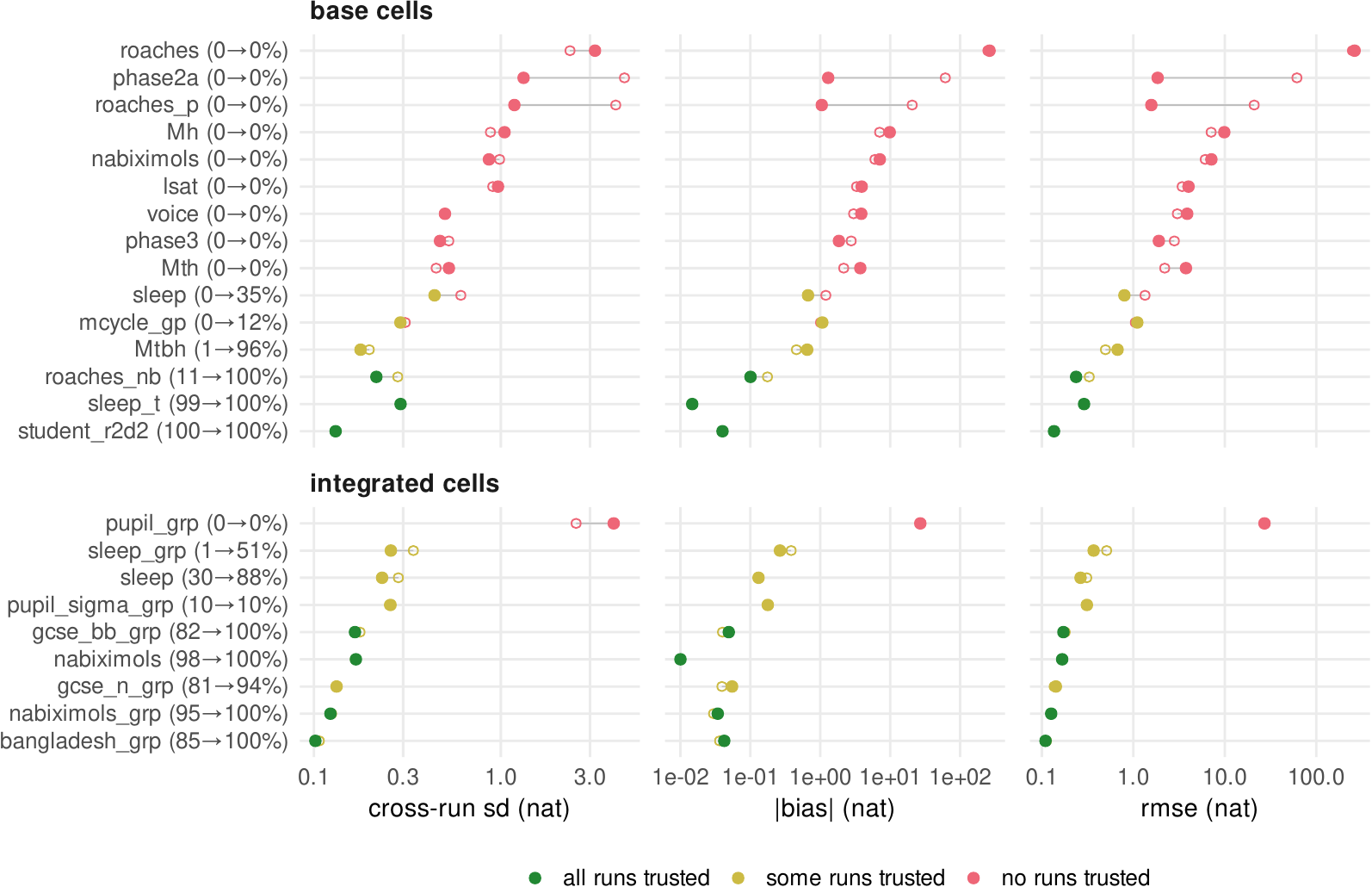}

}

\caption{\label{fig-psis-mm}Moment matching changes the PSIS estimate
little on most cells and makes it worse on some. PSIS and PSIS + MM per
cell: realized SD, absolute bias and RMSE, log axes; open points: plain
PSIS, filled points: PSIS + MM, each colored by the trusted \% of its
method; trusted \% before and after in parentheses.}

\end{figure}

Figure~\ref{fig-psis-mm} shows the realized SD, the absolute bias and
the realized RMSE of PSIS (open points) and PSIS + MM (filled points) on
log scales, joined by a gray line. The label of each cell gives the
trusted \% before and after matching. Moment matching does not reliably
improve the estimate. It moves the estimate by tens of nats on phase2a
(\(+61\) to \(+1.3\) nat) and roaches\_p (\(+21\) to \(+1.0\) nat), but
both cells stay untrusted. It makes the estimate worse on lsat (\(+3.3\)
to \(+3.9\) nat) and Mth (\(+2.2\) to \(+3.7\) nat), which agrees with
\textcite{Riha-Jedhoff-Buerkner-Vehtari:2026} and
\textcite{Bindoff:2026}. The bias predictor after matching is not
usable: its mean over the runs is above the realized bias on mcycle\_gp
(23 against 1.1 nat) and below it on roaches (27 against 263 nat).

\subsubsection{Is moment matching useful as a
check?}\label{is-moment-matching-useful-as-a-check}

Mostly yes. On roaches, voice, phase3, nabiximols, Mh and pupil\_grp
(int-an), 8\% or more of the folds have
\(\hat{k}_{\mathrm{psis}} > 0.7\) in almost every run, so the check is
not used there, and their runs stay untrusted. On roaches\_nb, Mtbh and
7 of the 9 integrated cells, matching raises the trusted \%. The plain
PSIS estimate there has a bias of at most 0.5 nat, so these runs are
rightly trusted. The higher the \(\hat{k}_{\mathrm{psis}}\) of a fold
before matching, the less often matching brings \(\hat{k}_w\) below 0.7.
The check also trusts some biased runs: on sleep, 26 of the 35 trusted
runs have a PSIS error above 1 nat, and on mcycle\_gp 11 of 12. These
errors are at most 2.6 nat. On both cells, the bias of PSIS is about 1
nat, and its predicted RMSE is too small.

\par\noindent\colorbox{black!6}{\begin{minipage}{\dimexpr\linewidth-2\fboxsep\relax}

\textbf{Answer to Q3b PSIS + MM.} Moment matching is not a reliable
estimator: it does not reliably reduce the error, and its bias cannot be
predicted. As a check, it adds correct trust on 9 of 24 cells, but it
also trusts biased runs on sleep and mcycle\_gp. It is cheap and
implemented in the loo package \autocite{Vehtari+etal:2026:loo}, so it
stays in the workflow of Section~\ref{sec-workflow}. Where it is not
available, it can be skipped.

\end{minipage}}\par

\subsection{Q3c: How good is MixIS?}\label{sec-fast-mixis}

\subsubsection{How accurate is MixIS?}\label{how-accurate-is-mixis}

MixIS is accurate on ten LOO cells (sleep\_t, student\_r2d2,
roaches\_nb, phase3, voice, mcycle\_gp, Mh, Mtbh, Mth and lsat), within
0.5 nat of the reference, and on every integrated cell except
pupil\_grp, within 0.8 nat. It fails in two scenarios.
\emph{Concentrated fold weights:} for a few folds, the fold weights
\(v_{i,s}\) concentrate on a few draws, and the upward bias of the \(v\)
term is large (Section~\ref{sec-framework-diag}). This is the case for
roaches, roaches\_p, phase2a and nabiximols. \emph{Dominant mixture
components:} when the folds differ strongly in \(p(y_i \mid y_{-i})\), a
few components dominate the mixture, the importance sampling ESS of the
common weights falls, and the common term is no longer reliable
(Section~\ref{sec-balanced-mixis}). This is the case for every LOGO base
cell and for sleep.

\subsubsection{Does its self-diagnostic flag when MixIS is
wrong?}\label{does-its-self-diagnostic-flag-when-mixis-is-wrong}

\begin{figure}[t]

\centering{

\includegraphics[width=1\linewidth,height=\textheight,keepaspectratio]{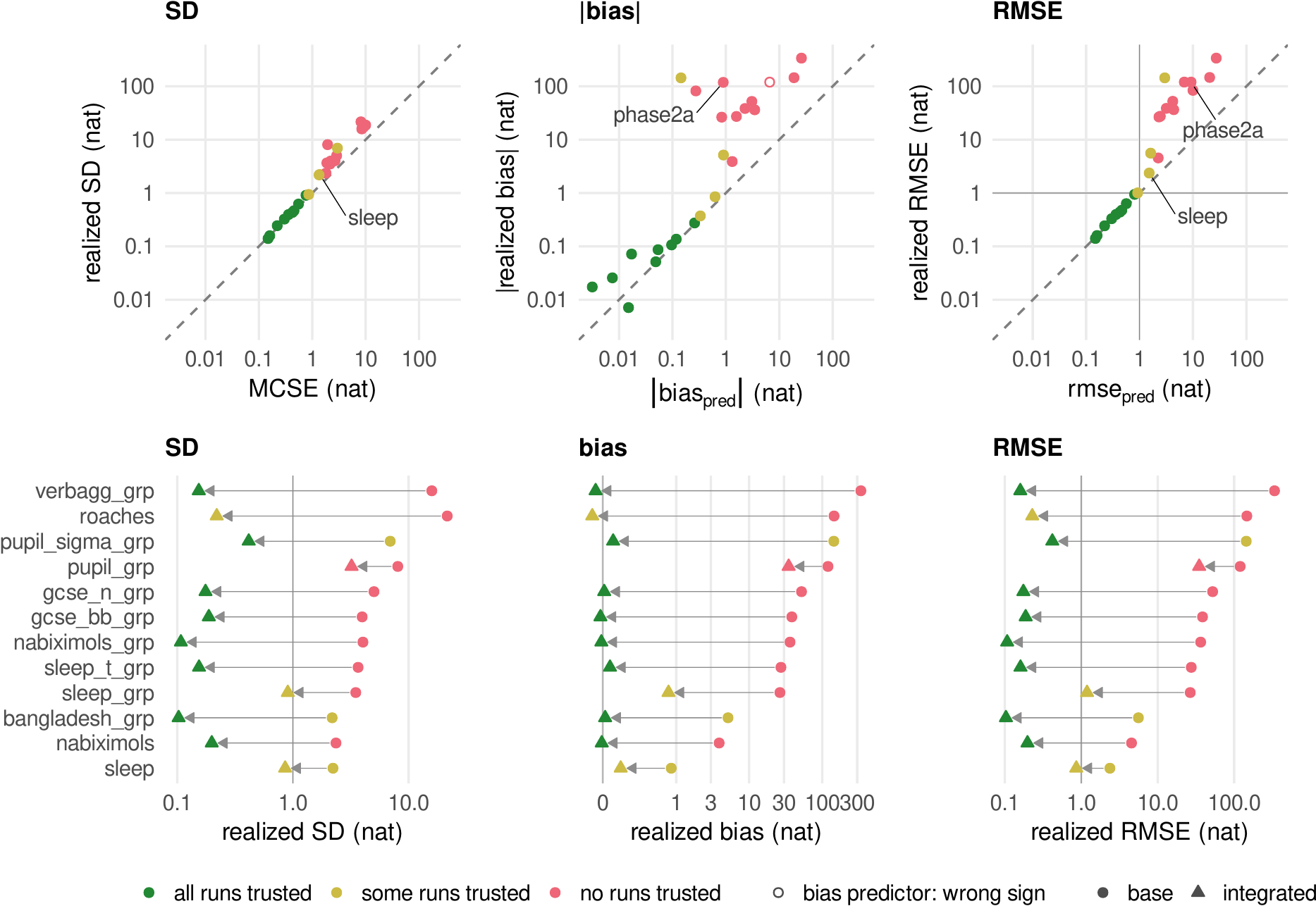}

}

\caption{\label{fig-mixis-calib}MixIS is calibrated on the base cells
where it is accurate and rejects almost every run where it fails, except
on sleep. MixIS at \(S = 4000\), colored by trusted \%. Top: predicted
against realized SD, absolute bias and RMSE, on one log scale, one point
per base cell; hollow: the bias predictor has the wrong sign; dashed
line: identity. Bottom: realized SD, bias and RMSE over the 100 runs for
each base cell (circle) and its integrated cell (triangle); the arrow
points from the base cell to the integrated cell; thin solid line: 1 nat
(SD, RMSE) and 0 (bias).}

\end{figure}

Figure~\ref{fig-mixis-calib} has the same layout as
Figure~\ref{fig-bf-calib}. Its trusted \% uses the trust rule alone,
without the convergence check of the mixture run
(Section~\ref{sec-framework-diag}). Almost every cell where MixIS fails
is red, and the trust rule rejects almost every run there: it trusts two
runs on bangladesh\_grp and one on pupil\_sigma\_grp. The predicted RMSE
is above 1 nat even where the bias predictor is far too small: on
phase2a, for example, the bias predictor gives \(+0.9\) nat against a
bias of \(+118\) nat, but the MCSE of 9.2 nat is enough to reject. Where
MixIS is accurate, 91--100\% of runs on the integrated cells are
trusted. The exception is sleep: its error exceeds 1 nat in 65\% of
runs, and the trust rule trusts 4 runs, three of them with an error
between 1.9 and 4.2 nat. On sleep\_grp (int), the predicted RMSE is half
of the realized RMSE (0.62 against 1.19 nat), and 44 of the 91 trusted
runs have an error above 1 nat.

\subsubsection{Does the self-diagnostic give the right sign and
magnitude of the
error?}\label{does-the-self-diagnostic-give-the-right-sign-and-magnitude-of-the-error-2}

Where MixIS is accurate, its total MCSE is close to the realized SD on
the base cells. These green cells lie on the diagonal in the top left
panel. On the accurate integrated cells, it is also close to the
realized SD, except on sleep, sleep\_grp, roaches and pupil\_sigma\_grp,
where it is 1.4 to 1.5 times too small. Counting the common term
correctly is essential: estimated separately, it would overstate the
total MCSE 15 to 20 times, and treating the folds as independent would
understate it up to three times. Where MixIS fails, the predicted error
is far too small: the red cells lie far above the diagonal in the top
right panel. Both intervals then rarely contain the error; on the
trusted cells, they contain it in about 95\% of the runs or somewhat
fewer. The bias predictor has the wrong sign on pupil\_grp (hollow
points in the top middle panel): it predicts about \(-7\) nat against a
realized \(+120\) nat.

\par\noindent\colorbox{black!6}{\begin{minipage}{\dimexpr\linewidth-2\fboxsep\relax}

\textbf{Answer to Q3c MixIS.} MixIS is accurate on more cells than PSIS
(RMSE below 1 nat on 18 against 14 of 35 cells), at the cost of one
additional MCMC run. Its self-diagnostic is calibrated on the base cells
where MixIS works and rejects almost every run where it fails, with
sleep as the main exception. Its failures on the LOGO cells come from a
few dominant components of the mixture.

\end{minipage}}\par

\begin{figure}[t]

\centering{

\includegraphics[width=1\linewidth,height=\textheight,keepaspectratio]{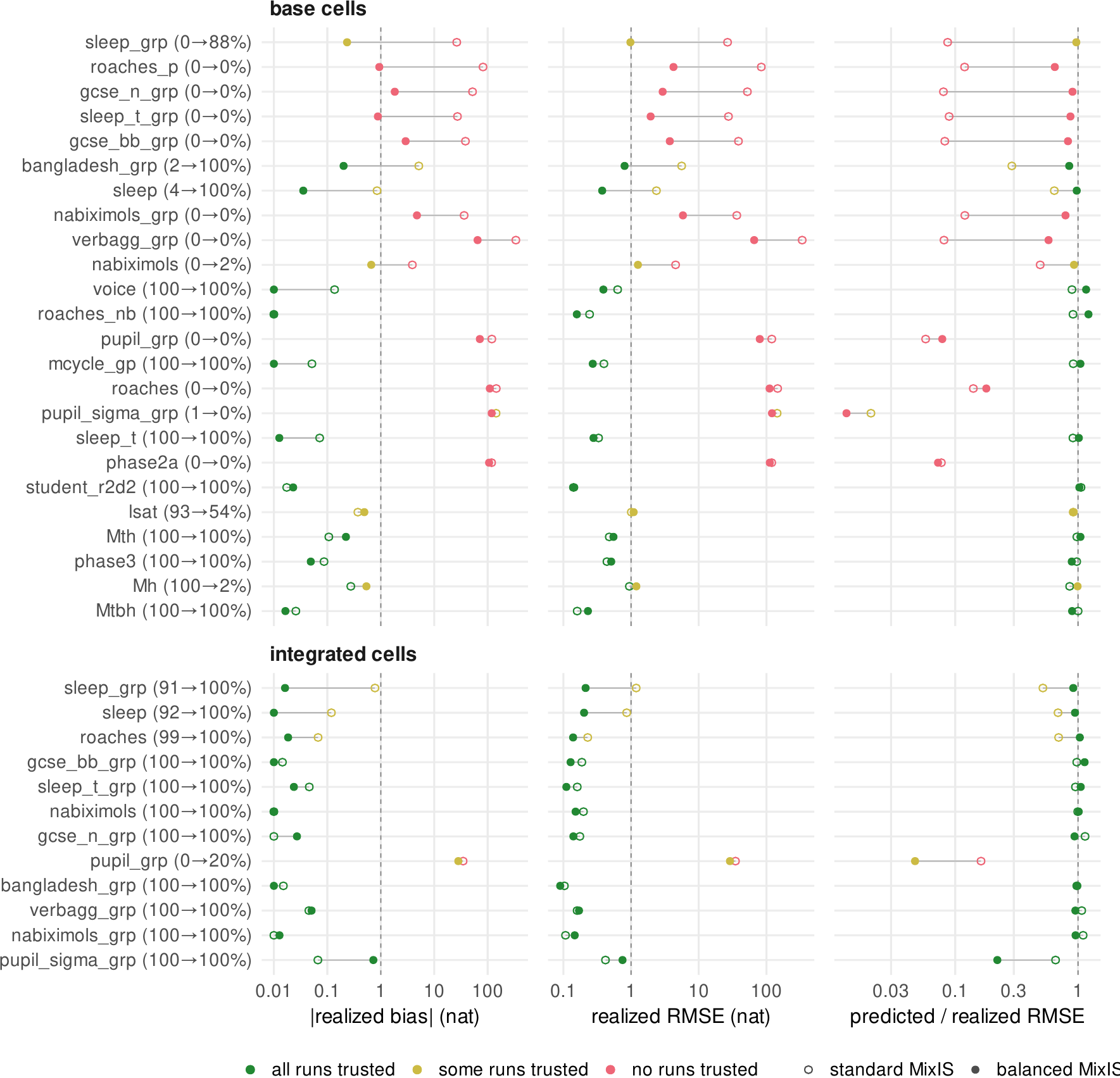}

}

\caption{\label{fig-bal-calib}Balancing reduces the RMSE on most cells
and moves the predicted RMSE toward the realized one. Open circle:
standard MixIS; filled: balanced MixIS; color: trusted \% of each method
(\(S = 4000\)). Left: absolute realized bias (values below 0.01 nat at
0.01); middle: realized RMSE; right: predicted / realized RMSE; dashed
line: 1. Base cells in the upper block, integrated cells in the lower
block.}

\end{figure}

\subsection{Q4: Does balancing reduce the error of
MixIS?}\label{sec-balanced-results}

bMixIS makes the folds share the mixture almost equally. On sleep, in
one run, the range of \(\pi_i\) over the folds changes from
\((6 \times 10^{-8}, 0.998)\) to \((0.0065, 0.0145)\). The error falls
from 4.4 to 0.5 nat, from above to below the 1-nat threshold.
Figure~\ref{fig-bal-calib} compares the two methods cell by cell. Its
realized RMSE is smaller than that of standard MixIS on 27 of the 36
cells computed with both. It also reduces the error on cells that stay
untrusted, for example on verbagg\_grp from 337 to 66 nat and on
roaches\_p from 84 to 4.2 nat. Balancing repairs the common term, but
not a data model that is too narrow, a fold with an extreme influence,
or many group-specific parameters. The bias therefore stays large on
phase2a (\(+106\) nat; too-narrow data model), on the two pupil cells
(\(+71\) and \(+119\) nat; extreme influence), and on the base cells of
roaches (\(+110\) nat) and verbagg\_grp (\(+65\) nat; many
group-specific parameters).

The five cells with a large remaining bias are red in
Figure~\ref{fig-bal-calib}: no run is trusted on them. On three cells
where standard MixIS fails, balancing makes the runs trusted: sleep,
sleep\_grp and bangladesh\_grp. The exception is pupil\_grp (int-an).
Its bias is \(+28.2\) nat, and the predicted RMSE alone trusts 20\% of
runs. The balanced mixture run does not converge in most runs. The
largest \(\widehat{R}\) \autocite{Vehtari-Gelman-Simpson-etal:2021} over
the parameters is above 1.05 in 71 of 100 runs, and \(\mathrm{ESS}(u)\)
is below 10 in 33. In the 12 runs with the largest \(\widehat{R}\) at
most 1.01, \(\hat{k}(u)\) is between 3.9 and 5.3. Their predicted RMSE
is between 0.7 and 1.7 nat, and their bias is as large as that of
standard MixIS. Every run that the predicted RMSE trusts has either
\(\widehat{R}\) above 1.05 or \(\hat{k}(u) > 0.7\). The likely cause is
the extreme influence of subject 20, but we have not verified this.

\begin{figure}[t]

\centering{

\includegraphics[width=1\linewidth,height=\textheight,keepaspectratio]{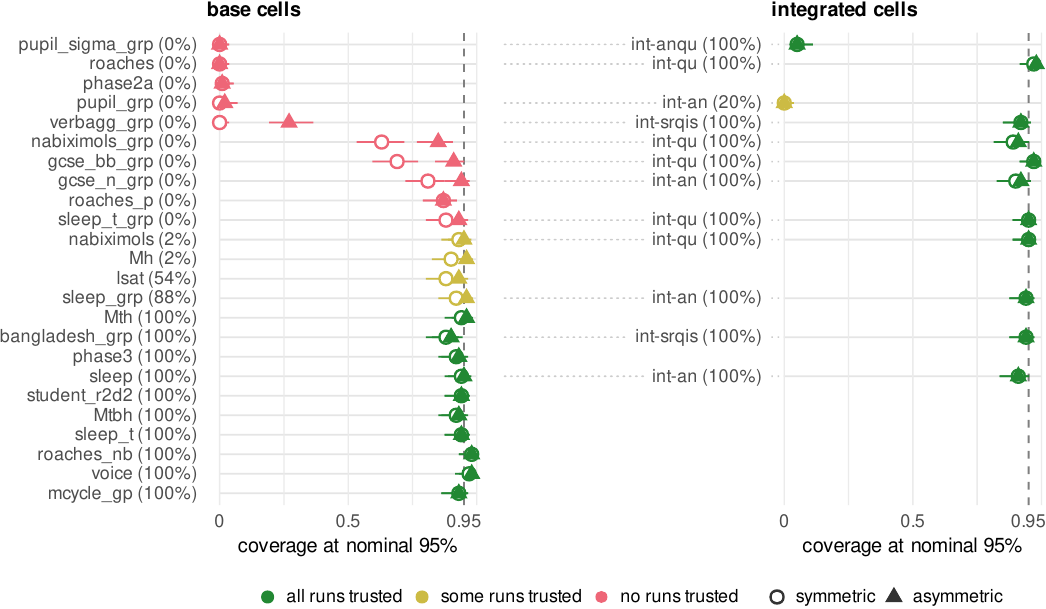}

}

\caption{\label{fig-bal-cover}The asymmetric interval of bMixIS contains
the error in close to 95\% of the runs, except where the bias is
structural. Share of runs in which the 95\% interval contains the error:
asymmetric interval (triangle) and symmetric interval (circle); dashed
line: 0.95; trusted \% after each cell name.}

\end{figure}

The self-diagnostic of bMixIS is better calibrated than that of standard
MixIS: its predicted RMSE is closer to the realized RMSE on 24 cells,
and its MCSE is closer to the realized SD on 25 cells. In the right
panel of Figure~\ref{fig-bal-calib}, most filled points lie closer to 1
than the open ones. Figure~\ref{fig-bal-cover} shows the coverage of the
two intervals, with the same markers as Figure~\ref{fig-bf-cover}. The
asymmetric interval contains the error in close to 95\% of the runs on
most cells. This holds also on the untrusted gcse and nabiximols\_grp
cells, where the bias predictor is close to the remaining bias.

Where the folds are already similar, bMixIS adds Monte Carlo variance.
On Mh and lsat, the trusted \% at \(S = 4000\) falls from 100\% to 2\%
and from 93\% to 54\%. This cost is finite-sample variance, which more
draws remove. With \(S = 16000\) the predicted RMSE falls from 1.19 to
0.61 nat on Mh and from 1.00 to 0.50 nat on lsat, and all runs are
trusted. The runs with \(S = 4000\) already predict this. Scaling the
MCSE as \(1/\sqrt{S}\) and the bias predictor as \(1/S\) gives 0.54 and
0.47 nat. So the cost of balancing can be predicted before more draws
are spent.

\par\noindent\colorbox{black!6}{\begin{minipage}{\dimexpr\linewidth-2\fboxsep\relax}

\textbf{Answer to Q4.} Balancing reduces the error of MixIS on 27 of 36
cells and improves the calibration of its predicted RMSE on 24, at the
same cost as MixIS. Where the folds are already similar, it adds Monte
Carlo variance, which more draws remove. We therefore balance by
default. Its trust rule fails on one cell, where the convergence check
or \(\hat{k}(u)\) warns in every trusted run.

\end{minipage}}\par

\section{Practical workflow}\label{sec-workflow}

The experiments show, for every method, when its self-diagnostic can be
trusted, but a user who needs one elpd must still decide which method to
run first and what to do when a run is not trusted. Based on the results
of our experiments, we combine the methods into a workflow that starts
with the cheapest method and moves to a more expensive one only when the
self-diagnostic of the cheaper method does not trust the run. The cost
then stays low where the problem is easy. The workflow extends the
stepwise algorithm of \textcite{Paananen+etal:2021:implicit}, which
moves from PSIS to MM to refitting according to Pareto-\(\hat{k}\), and
the practice of refitting only the folds with high Pareto-\(\hat{k}\)
\autocite{Vehtari+etal:PSIS:2024}. The difference is that every step now
has a self-diagnostic for the total elpd, not only a check of each fold.

We recommend the following steps:

\begin{enumerate}
\def\labelenumi{\arabic{enumi}.}
\tightlist
\item
  Run MCMC for the full data with the model as written, with the
  group-specific parameters sampled.
\item
  Compute PSIS (Section~\ref{sec-fast-psis}). Stop if every fold has
  \(\hat{k}_{\mathrm{psis}} \leq 0.7\) and the predicted RMSE is below 1
  nat. On our cells this happened rarely, but they were chosen to be
  difficult; in practice it is common.
\item
  If only a few folds have \(\hat{k}_{\mathrm{psis}} > 0.7\)
  (Section~\ref{sec-framework-mm}), use MM as a check of the PSIS
  estimate (Section~\ref{sec-fast-mm}), and stop if the check trusts the
  run. This check is cheap and keeps further computation simple, but it
  does not detect every biased run. In our experiments, the errors of
  the wrongly trusted runs were at most 2.6 nat. Where MM is not
  available, skip this step.
\item
  Run bMixIS (Section~\ref{sec-balanced-mixis}), with the PSIS estimates
  as balancing weights, and keep the mixture draws. Check the
  convergence of the mixture run, including \(\mathrm{ESS}(u)\), and
  stop if the run is trusted.
\item
  If the model has group-specific parameters and integrating them out is
  feasible, integrate them out from the saved draws
  (Section~\ref{sec-integration}), and stop if integrated PSIS or
  integrated bMixIS is trusted. Where a good integration rule is already
  known, this step can come before step 4. Where the Laplace
  approximation is used, use the importance sampling correction.
\item
  Run brute-force and the CV bridge together
  (Section~\ref{sec-reference-bridge}), with integration if it is
  available. The two use the same draws, so both results come from one
  computation, and the draws of each fold need not be stored. This step
  is expensive, but it is the most likely to be accurate, and its
  reliability checks show the folds where it is not.
\end{enumerate}

An accurate elpd is not always needed. The roaches data show why: with a
Poisson data model (roaches\_p) brute-force, PSIS and MixIS all fail,
while with a negative-binomial model (roaches\_nb) every method is
within about 1 nat of the reference (Section~\ref{sec-experiments}). If
the computational problems come from a misspecified model, and
predictive checks already show a better model, there is little reason to
compute the elpd of the misspecified model accurately. The same holds
when two models differ in elpd by much more than the predicted RMSE of
an inexpensive method. The self-diagnostic makes this decision possible,
because it gives the magnitude of the computational error, which the
user can compare with the error they would accept.

\section{Discussion}\label{sec-discussion}

In this paper, we investigated the Monte Carlo error of Bayesian CV
methods. We checked how well current diagnostics flag the cases where a
CV estimate is wrong, and where they fall short. We then gave every CV
method a self-diagnostic: an estimate of the error of the total CV
estimate, together with a trust rule that tells a user whether to trust
the estimate. The trust rule combines the predicted RMSE, which joins
the MCSE and the bias predictor, with the reliability checks of each
method, such as Pareto-\(\hat{k}\) for PSIS or the overlap check of the
CV bridge, into one decision.

Earlier evaluations of CV approximations used brute-force as the
reference and treated it as exact
\autocite{Vehtari+etal:2016:LOO_for_GLVM,Vehtari+etal:PSIS-LOO:2017,Silva+Zanella:2024:robustloo,Bindoff:2026}.
Our results do not invalidate these evaluations, but they narrow the
conditions under which they hold. When the data model fits the data and
no fold has an extreme influence, brute-force is accurate on 18 of the
19 base cells (Section~\ref{sec-bruteforce}). It fails, however, in
three scenarios, and its trust rule rejects every run in all three.
\emph{Too-narrow data model:} a few observations lie far in the tail of
their predictive distribution, as on roaches\_p (\(-17\) nat) and
phase2a (\(-78\) nat). Such a failure is then also a sign that the model
does not fit, and the remedy may be a better model rather than a better
CV method. \emph{Many folds:} small fold biases add up, as on
verbagg\_grp with 316 folds (\(-5\) nat). \emph{Influential groups:}
when a left-out group has many observations, few draws from its
population distribution fall where its data are likely. This is the case
for the two pupil cells, with up to 120 observations per group and
errors of about \(-48\) nat. Integration removes the whole error on
pupil\_sigma\_grp, but on pupil\_grp it removes less than half (\(-48\)
to \(-30\) nat).

The self-diagnostic of brute-force also explains why
\textcite{Paananen+etal:2021:implicit} found Pareto-\(\hat{k}\) to flag
biased runs, while \textcite{Riha-Jedhoff-Buerkner-Vehtari:2026}
observed accurate estimates with \(\hat{k}\) above 0.7
(Section~\ref{sec-background-bf}). Pareto-\(\hat{k}\) sees that the
values \(p(y_i \mid \theta_s)\) have a heavy tail, but not where this
tail ends. The truncGPD MCSE adds a largest plausible value,
\(p_{\max}\), and turns the shape of the tail into an error in nat
(Section~\ref{sec-diag-bf}).

The CV bridge is a good general reference: it is accurate enough to
serve as the reference on 21 of our 24 base cells
(Table~\ref{tbl-cells}). On roaches, it carries an offset of about
\(-0.15\) nat against integrated PSIS, which we therefore use as the
reference. On pupil\_grp, its checks correctly reject the folds where it
fails, in every run. The CV bridge is more accurate than brute-force for
two reasons. It uses the full-data draws, which brute-force ignores, in
addition to the leave-fold-out draws. And its terms are bounded, so a
few extreme draws cannot dominate the estimate, as they do for
brute-force when \(p(y_i \mid \theta_s)\) has a heavy tail. It is also
more accurate than bridge sampling of the two marginal likelihoods
\autocite{Riha-Jedhoff-Buerkner-Vehtari:2026}, because it estimates
their ratio directly, so that much of the error cancels.

We checked the CV bridge in two ways
(Section~\ref{sec-reference-check}). From \(S = 4000\) to \(S = 16000\),
its mean changes by at most 0.13 nat on 23 of the 24 references and by
0.31 nat on verbagg\_grp, so a bias that shrinks with \(S\) is small.
Integrated PSIS, which shares no computation with the CV bridge, agrees
with the references to within 0.38 nat on every integrated cell except
roaches, whose reference it is, and pupil\_grp, where it fails. A bias
that does not shrink with \(S\) is therefore unlikely on these cells.

PSIS is accurate when every \(\hat{k}_{\mathrm{psis}}\) is at most 0.7,
the threshold of \textcite{Vehtari+etal:PSIS:2024}. Above it, PSIS is
usually biased, and its trust rule rejects the runs where some fold
exceeds it, also the few accurate ones. Moment matching does not
reliably improve the PSIS estimate, in agreement with
\textcite{Riha-Jedhoff-Buerkner-Vehtari:2026} and
\textcite{Bindoff:2026}. It often brings \(\hat{k}\) below 0.7 without
improving the estimate, probably because the transform is fitted to the
same draws that it then reweights
\autocite{Riha-Jedhoff-Buerkner-Vehtari:2026}. Fitting the transform on
one half of the draws and the estimate on the other half removes this
bias, but the variance grows so much that the RMSE does not decrease
(Code availability, section \emph{PSIS + moment matching}). Moment
matching is therefore useful as a check on PSIS rather than as an
estimator. As a check it adds correct trust on 9 of 24 cells, but on
sleep and mcycle\_gp it also trusts biased runs.

MixIS is more robust than PSIS \autocite{Silva+Zanella:2024:robustloo},
but \textcite{Silva+Zanella:2024:robustloo} gave no MCSE or diagnostic,
so a user could not decide whether to trust an estimate. With the
self-diagnostic, this decision becomes possible: MixIS is more accurate
than PSIS on most cells, and its trust rule rejects almost every run
where it fails, with sleep as the main exception. The false trust on
sleep comes from the common term of the mixture: where a few components
dominate, it is poorly estimated, and the predicted RMSE is too small.
\textcite{Silva+Zanella:2024:robustloo} also noted that unequal weights
of the mixture components might improve efficiency, but they gave no
rule for choosing them. We set the weights proportional to
\(\exp(\widehat{\mathrm{elpd}}_{\mathrm{psis},i})\), which balances the
mixture across the folds. Balancing reduces the error on 27 of the 36
cells and repairs the failure of MixIS on sleep, sleep\_grp and
bangladesh\_grp, but it cannot repair a data model that is too narrow, a
fold with an extreme influence, or the effect of many group-specific
parameters. Its trust rule fails on one cell, pupil\_grp (int-an), where
the convergence check or \(\hat{k}(u)\) warns in every trusted run.
Where the folds are already similar, balancing adds Monte Carlo
variance, but more draws remove it, and the self-diagnostic of a run
with \(S = 4000\) predicts this, as on Mh and lsat.

Integration is not a CV method but a change to what is computed, so it
can be combined with any of them (Section~\ref{sec-integration}). It
reduced the error of every method on most LOGO cells. For PSIS, it
turned biases of \(+7\) to \(+340\) nat on the LOGO base cells into at
most 0.38 nat on their integrated cells, except on pupil\_grp
(Section~\ref{sec-fast-psis}). Its cost depends on the rule. Integrating
within MCMC is faster for the analytic and mode-centered Gauss--Hermite
marginals, and integrating after MCMC is faster for the Laplace and the
other quadrature marginals (Code availability, appendix
\emph{Integration rules per cell and timing}). After MCMC, the integral
can also be computed only when a self-diagnostic warns.

The main risk of integration is that the self-diagnostic cannot see the
error of the integral, because the integral is computed in the same way
in every draw and every run. With the plain Laplace approximation,
integrated PSIS on verbagg\_grp has a bias of \(-22\) nat, almost the
same in every fold, yet its largest \(\hat{k}_{\mathrm{psis}}\) is at
most 0.44, its MCSE is close to the realized SD, and every run is
trusted. Integrated MixIS with the same approximation has the same bias,
so the agreement between the two methods says nothing about the integral
that they share. The importance sampling correction removes the error of
the Laplace approximation. A quadrature rule can fail in the same way:
with a 64-node Gauss--Hermite rule placed according to the population
distribution, the kind used by \textcite{Bindoff:2026}, integrated PSIS,
MixIS and balanced MixIS are 4.5--7 nat below the reference on
pupil\_sigma\_grp, again with normal diagnostics, whereas a
mode-centered rule is within a few tenths of a nat (Code availability,
appendix \emph{An independent integration route is needed to check the
marginal}). Quadrature rules can be tested by increasing the number of
nodes, and checking whether the result eventually stabilizes (this is
the usual adaptive quadrature approach).

\subsection{Limitations}\label{sec-limitations}

The thresholds of the trust rules are choices: 0.7 for
Pareto-\(\hat{k}\) and 0.3 for the bridge MCSE follow earlier work
\autocite{Vehtari+etal:PSIS:2024,Micaletto+Vehtari:2025}, the 1-nat
threshold and the others are ours, and none is derived from a bound. The
probability 0.95 of the truncation point \(\tau_M\) is also a choice:
when \(\tau_M\) sets the truncation, the predicted errors are smaller
with 0.90 and larger with 0.99. The 1-nat threshold already causes false
positives, as on gcse\_n\_grp and sleep\_grp, where brute-force is
accurate but almost every run is rejected, and it does not depend on the
difference between the models being compared: a user whose models differ
by 20 nat can accept a larger error than one whose models differ by 2
nat. Folds with an extreme influence remain the hardest cases, as on
pupil\_grp. \textcite{Bindoff:2026} predicts such failures of PSIS from
the structure of the model, before any computation, but we know of no
similar prediction for brute-force and the mixture methods. We studied
LOO and LOGO with folds that are modelled as conditionally independent
given the parameters, and the Monte Carlo error of a single elpd, not
the error of a difference between two models or the uncertainty from
having finite data \autocite{Sivula-Magnusson-Vehtari:2025}. All runs
used the No-U-Turn sampler of Stan with Pathfinder initialization, and
the mixture methods may behave differently with other samplers. We did
not compare the computing time of the CV methods. Finally, our cells
were chosen to make the methods fail, so the shares in this paper are
not estimates of how often each method fails in practice.

\section*{Code availability}\label{code-availability}
\addcontentsline{toc}{section}{Code availability}

The case-study notebook \emph{Cross-validation diagnostics}
(\url{https://users.aalto.fi/~ave/casestudies/cv_diagnostics/cv_diagnostics.html})
is the supplementary material of this paper. It gives the derivations,
the per-cell results of every method and the code that produces the
figures, from the results of the R package loobench
(\url{https://github.com/avehtari/loobench}), which implements every
method of this paper. The notebook uses some earlier names for the same
quantities.

\section{Acknowledgements}\label{acknowledgements}

The authors thank Jonah Gabry, Seth Axen and Anna Riha for valuable
feedback on the manuscript. This work was supported by the Research
Council of Finland grants (368069 and 372494).

\section*{References}\label{references}
\addcontentsline{toc}{section}{References}

\printbibliography[heading=none]

\appendix

\section*{Appendix}\label{appendix}
\addcontentsline{toc}{section}{Appendix}

\setcounter{subsection}{0}
\renewcommand{\thesubsection}{\Alph{subsection}}

\subsection{Notation and abbreviations}\label{sec-notation}

\begin{longtable}[]{@{}
  >{\raggedright\arraybackslash}p{(\linewidth - 4\tabcolsep) * \real{0.2533}}
  >{\raggedright\arraybackslash}p{(\linewidth - 4\tabcolsep) * \real{0.5733}}
  >{\raggedright\arraybackslash}p{(\linewidth - 4\tabcolsep) * \real{0.1733}}@{}}
\caption{Notation and abbreviations, with the section where each is
defined. }\label{tbl-notation}\tabularnewline
\toprule\noalign{}
\begin{minipage}[b]{\linewidth}\raggedright
Symbol or term
\end{minipage} & \begin{minipage}[b]{\linewidth}\raggedright
Meaning
\end{minipage} & \begin{minipage}[b]{\linewidth}\raggedright
Defined in
\end{minipage} \\
\midrule\noalign{}
\endfirsthead
\toprule\noalign{}
\begin{minipage}[b]{\linewidth}\raggedright
Symbol or term
\end{minipage} & \begin{minipage}[b]{\linewidth}\raggedright
Meaning
\end{minipage} & \begin{minipage}[b]{\linewidth}\raggedright
Defined in
\end{minipage} \\
\midrule\noalign{}
\endhead
\bottomrule\noalign{}
\endlastfoot
CV; LOO, LOGO & cross-validation; leave-one-out, leave-one-group-out &
Section~\ref{sec-intro}, Section~\ref{sec-background-elpd} \\
elpd, \(\mathrm{elpd}_i\) & sum over folds of
\(\log p(y_i \mid y_{-i})\); fold term &
Section~\ref{sec-background-elpd} \\
\(y\); \(y_i\), \(y_{-i}\) & data; left-out observation and the data
without it & Section~\ref{sec-background-elpd} \\
\(i\), \(n\) & fold index; number of folds (LOO) &
Section~\ref{sec-background-elpd} \\
\(g\), \(y_g\), \(y_{-g}\) & group index (LOGO); data of group \(g\) and
the data without it & Section~\ref{sec-background-elpd} \\
\(\theta\); \(b_g\), \(\phi\) & all parameters; group-specific and
shared parameters & Section~\ref{sec-background-elpd} \\
\(p(\theta \mid y_{-i})\) & leave-fold-out posterior &
Section~\ref{sec-background-elpd} \\
\(p(y_g \mid \phi)\) & likelihood of group \(g\) with \(b_g\) integrated
out & Section~\ref{sec-background-elpd} \\
\(S\) & number of posterior draws per run (4000; 16000 for the
references) & Section~\ref{sec-background-bf},
Section~\ref{sec-testbed-cells} \\
\(s\), \(\theta_s\) & draw index; draw \(s\) &
Section~\ref{sec-background-bf} \\
\(\hat{p}_i\) & a method's estimate of \(p(y_i \mid y_{-i})\) &
Section~\ref{sec-background-bf} \\
\(a_{i,s}\), \(\bar{a}_i\) & values computed from draw \(s\) for fold
\(i\); their Monte Carlo mean & Section~\ref{sec-framework-diag} \\
MCSE; \(\mathrm{mcse}_i\); \(\mathrm{mcse}_u\), \(\mathrm{mcse}_{v,i}\)
& Monte Carlo standard error; MCSE of fold term \(i\); MCSEs of the two
MixIS terms & Section~\ref{sec-background-bf},
Section~\ref{sec-framework-diag}, Section~\ref{sec-diag-mixis} \\
\(w_{i,s}\), \(\tilde{w}_{i,s}\) & importance weight; smoothed weight
scaled to sum to one & Section~\ref{sec-background-psis},
Section~\ref{sec-diag-psis} \\
GPD; \(M\); \(c\), \(\hat{\sigma}\) & generalized Pareto distribution;
number of largest values in the tail fit; its threshold and scale &
Section~\ref{sec-background-psis}, Section~\ref{sec-diag-bf} \\
Pareto-\(\hat{k}\) & shape of the generalized Pareto tail fit;
\(\hat{k}_{\mathrm{psis}}\) (PSIS weights), \(\hat{k}_{\mathrm{bf}}\)
(brute-force values), \(\hat{k}_{\mathrm{mm}}\), \(\hat{k}_w\) (after
moment matching), \(\hat{k}(u)\) (MixIS common weights) &
Section~\ref{sec-background-psis}, Section~\ref{sec-diag-psis},
Section~\ref{sec-framework-mm}, Section~\ref{sec-diag-mixis} \\
MM & moment matching & Section~\ref{sec-background-mm} \\
MixIS; \(q_{\mathrm{mix}}\) & mixture importance sampling; its mixture
proposal & Section~\ref{sec-background-mixis} \\
\(u_s\), \(v_{i,s}\) & common weights and fold weights of MixIS &
Section~\ref{sec-background-mixis}, Section~\ref{sec-diag-mixis} \\
\(q_1\), \(q_2\); \(Z_1\), \(Z_2\); \(h\) & unnormalized densities;
their normalizing constants; bridge function &
Section~\ref{sec-background-bridge} \\
conditional, integrated & computation that samples \(b_g\) / integrates
it out & Section~\ref{sec-integration} \\
run; \(r\), \(R\) & one computation with one set of draws; run index;
100 runs & Section~\ref{sec-background-bf},
Section~\ref{sec-testbed-cells} \\
\(m\); \(\widehat{\mathrm{elpd}}_m\), \(\widehat{\mathrm{elpd}}_{m,i}\)
& method; its estimate of the elpd and of fold term \(i\) in one run &
Section~\ref{sec-framework-diag} \\
subscript bf, psis, mixis, bmixis, bridge & quantity of that method,
e.g.~\(\widehat{\mathrm{elpd}}_{\mathrm{bf},i}\),
\(\mathrm{mcse}_{\mathrm{psis},i}\) &
Section~\ref{sec-framework-diag} \\
self-diagnostic; reliability checks; trust rule & error estimate with
its checks; tests of whether the error estimate can be believed;
decision trusted or not trusted & Section~\ref{sec-intro},
Section~\ref{sec-framework-diag} \\
\(\mathrm{mcse}_{\mathrm{total}}\) & MCSE of the elpd estimate &
Section~\ref{sec-framework-diag} \\
\(\mathrm{bias}_{\mathrm{pred}}\); \(\mathrm{rmse}_{\mathrm{pred}}\) &
bias predictor; predicted RMSE & Section~\ref{sec-framework-diag} \\
asymmetric interval & 95\% interval for the error, wider on the side of
\(\mathrm{bias}_{\mathrm{pred}}\) & Section~\ref{sec-framework-diag} \\
IJ; \(\psi_{i,s}\), \(\psi_s\); \(\psi^{\mathrm{f}}_{i,s}\),
\(\psi^{\mathrm{l}}_{i,s}\) & infinitesimal jackknife; influence of draw
\(s\) on fold \(i\) and on the total; influences of the full-data and
leave-fold-out draws in the CV bridge &
Section~\ref{sec-framework-diag}, Section~\ref{sec-methods-new} \\
RSS & root sum of squares of the fold MCSEs &
Section~\ref{sec-framework-diag} \\
ESS, relESS; \(\mathrm{relESS}_i\) & effective sample size; ESS divided
by \(S\); relESS of the influences of fold \(i\) &
Section~\ref{sec-framework-diag}, Section~\ref{sec-diag-psis} \\
truncGPD; \(p_{\max}\); \(\tau\), \(\tau_M\), \(\tau_{\mathrm{eff}}\) &
truncated generalized Pareto MCSE; largest plausible value of
\(p(y_i \mid \theta_s)\); truncation points &
Section~\ref{sec-diag-bf} \\
\(q_{\mathrm{full}}\), \(q_{\mathrm{loo}}\); \(\hat{\lambda}_i\);
\(\theta^{\mathrm{f}}_s\), \(\theta^{\mathrm{l}}_s\);
\(\omega_{\mathrm{f}}\), \(\omega_{\mathrm{l}}\); \(N_{i,s}\),
\(D_{i,s}\) & unnormalized full-data and leave-fold-out posteriors; CV
bridge estimate of \(p(y_i \mid y_{-i})\); full-data and leave-fold-out
draws; ESS shares; terms of the estimate &
Section~\ref{sec-methods-new} \\
cross-density gap \(g_i\); MM-mean; \(\tilde{\theta}^{\mathrm{f}}_s\),
\(\tilde{\lambda}_i\) & overlap measure of the CV bridge; shift of the
full-data draws; shifted draws and the estimate computed with them &
Section~\ref{sec-methods-new} \\
\(\mathrm{ESS}(u)\) & MCMC ESS of the common weights of MixIS &
Section~\ref{sec-diag-mixis} \\
\(\mathrm{ESS}_{\mathrm{IS}}(u)\),
\(\mathrm{relESS}_{\mathrm{IS}}(\alpha)\) & importance sampling ESS of
the common weights; its prediction divided by \(S\) &
Section~\ref{sec-balanced-mixis} \\
\(\alpha_i\); \(\pi_i\); \(m_\alpha\) & balancing weights; mixture share
of fold \(i\); balanced mixture sum & Section~\ref{sec-methods-new} \\
\(L\), \(J\) & points and randomizations of the importance sampling
correction of the Laplace approximation & Appendix \ref{sec-cells} \\
cell; base cell, integrated cell & one model and data set; the model
with \(b_g\) sampled, and with \(b_g\) integrated out &
Section~\ref{sec-testbed-cells} \\
\(\mathrm{err}_r\); \(\mathrm{elpd}_{\mathrm{ref}}\) & error of run
\(r\); reference value (mean of 100 reference runs) &
Section~\ref{sec-testbed-cells} \\
realized bias, SD, RMSE; coverage; trusted \% & over 100 runs against
the reference; share of runs with \(\mathrm{err}_r\) inside an interval;
share of trusted runs & Section~\ref{sec-testbed-cells} \\
\end{longtable}

\subsection{The cells of the experiments}\label{sec-cells}

Table~\ref{tbl-cells} lists the 37 cells of
Section~\ref{sec-experiments}, which Section~\ref{sec-testbed-cells}
describes. This section also explains why we selected each cell.

The cells phase2a and phase3 were used by
\textcite{Paananen+etal:2021:implicit} to illustrate the benefits of
moment matching. The cell voice was used by
\textcite{Silva+Zanella:2024:robustloo} to illustrate the benefits of
mixture importance sampling. The cell roaches is an example where PSIS,
PSIS+MM and MixIS fail. The cell roaches\_nb is an easier variant, and
roaches\_p is a harder misspecified variant. The cell nabiximols is a
hierarchical model with a non-Gaussian data model and a small number of
observations per group, and nabiximols\_grp is the LOGO variant. The
cell sleep is a Gaussian hierarchical model with a known model
misspecification, and sleep\_t is the corresponding variant with a more
robust data model. The cells sleep\_grp and sleep\_t\_grp are the LOGO
variants. The cell student\_r2d2 is an example of a regression with a
moderately high-dimensional posterior that is still easy for PSIS. The
cells roaches, roaches\_nb, roaches\_p, nabiximols, sleep, sleep\_t and
student\_r2d2 were used by \textcite{Gelman-Vehtari-Mcelreath-etal:2026}
to demonstrate PSIS and model comparison. The cell sleep\_grp was used
by \textcite{Riha-Jedhoff-Buerkner-Vehtari:2026} in LOGO experiments.

The cells Mh, Mtbh and Mth differ from the other cells because they are
capture--recapture models, and they also have quite high-dimensional
posteriors. The cell lsat has a high-dimensional posterior and is known
to be difficult for marginal likelihood estimation with bridge sampling
\autocite{Micaletto+Vehtari:2025}. The cell mcycle\_gp is a flexible
Gaussian process model with a different type of posterior dependencies
than plain hierarchical models. The cells Mh, Mtbh, Mth, lsat and
mcycle\_gp are from posteriordb
\autocite{Magnusson-etal:2025:posteriordb} and were used for bridge
sampling experiments on the marginal likelihood by
\textcite{Micaletto+Vehtari:2025}.

The cell bangladesh\_grp is a non-Gaussian hierarchical model previously
used by \textcite{McElreath:2020}. The pupil data set has many
observations per group (median 120), which makes many groups highly
influential in LOGO. The cells gcse\_n\_grp and pupil\_grp use a normal
model that allows analytic integration, and gcse\_bb\_grp and
pupil\_sigma\_grp are more challenging variants that require quadrature
integration. The pupil data were used by
\textcite{Nicenboim-Schad-Vasishth:2025} in exercises on hierarchical
models. The gcse data (data set Gcsemv in the mlmRev package,
\textcite{Bates-Maechler-Bolker:2026}) were used by
\textcite{Lee-Sim-Ji-etal:2018} in a tutorial on multilevel models.
Variants of the verbagg\_group model were used by
\textcite{Merkle+Furr+Rabe-Hesketh:2019} and
\textcite{Riha-Jedhoff-Buerkner-Vehtari:2026} in LOGO experiments.

\begin{longtable}[]{@{}
  >{\raggedright\arraybackslash}p{(\linewidth - 10\tabcolsep) * \real{0.0897}}
  >{\raggedright\arraybackslash}p{(\linewidth - 10\tabcolsep) * \real{0.3205}}
  >{\raggedright\arraybackslash}p{(\linewidth - 10\tabcolsep) * \real{0.2564}}
  >{\raggedleft\arraybackslash}p{(\linewidth - 10\tabcolsep) * \real{0.0641}}
  >{\raggedleft\arraybackslash}p{(\linewidth - 10\tabcolsep) * \real{0.0641}}
  >{\raggedright\arraybackslash}p{(\linewidth - 10\tabcolsep) * \real{0.2051}}@{}}
\caption{The 37 cells. A label in parentheses adds the integrated cell
of the same model. Dim. is the dimension of the posterior. Reference is
the method that gives the reference value
(Section~\ref{sec-reference-check}); ``+ MM'' means that the MM-mean of
Section~\ref{sec-methods-new} was used on at least one fold, and
``(int.)'' that the reference is computed on the integrated model.
Deviations from the rule of Section~\ref{sec-testbed-cells}: phase2a
uses the MM-mean on every fold, because it is more precise there than
the trust rule (RMSE 0.04 against 0.38 nat at \(S = 16000\)); the gap
check rejects one fold in every run; roaches\_p uses the plain estimate
on every fold, because at \(S = 16000\) the MM-mean is used on only
0.15\% of the folds and changes the reference by only 0.003 nat; phase3
uses the plain estimate on every fold, because it passes the first
accuracy check of Section~\ref{sec-reference-check} without the MM-mean;
roaches uses integrated PSIS, because the CV bridge differs from it by
about \(-0.15\) nat even with the MM-mean; on the pupil cells the two
posteriors of the conditional model overlap on only 1 or 2 of the 20
folds, so their references are computed on the integrated model.
}\label{tbl-cells}\tabularnewline
\toprule\noalign{}
\begin{minipage}[b]{\linewidth}\raggedright
Scope
\end{minipage} & \begin{minipage}[b]{\linewidth}\raggedright
Cell
\end{minipage} & \begin{minipage}[b]{\linewidth}\raggedright
Data model
\end{minipage} & \begin{minipage}[b]{\linewidth}\raggedleft
Folds
\end{minipage} & \begin{minipage}[b]{\linewidth}\raggedleft
Dim.
\end{minipage} & \begin{minipage}[b]{\linewidth}\raggedright
Reference
\end{minipage} \\
\midrule\noalign{}
\endfirsthead
\toprule\noalign{}
\begin{minipage}[b]{\linewidth}\raggedright
Scope
\end{minipage} & \begin{minipage}[b]{\linewidth}\raggedright
Cell
\end{minipage} & \begin{minipage}[b]{\linewidth}\raggedright
Data model
\end{minipage} & \begin{minipage}[b]{\linewidth}\raggedleft
Folds
\end{minipage} & \begin{minipage}[b]{\linewidth}\raggedleft
Dim.
\end{minipage} & \begin{minipage}[b]{\linewidth}\raggedright
Reference
\end{minipage} \\
\midrule\noalign{}
\endhead
\bottomrule\noalign{}
\endlastfoot
LOO & Mh & capture--recapture & 385 & 388 & CV bridge \\
LOO & Mtbh & capture--recapture & 146 & 154 & CV bridge \\
LOO & Mth & capture--recapture & 387 & 394 & CV bridge \\
LOO & lsat & Rasch & 1000 & 1006 & CV bridge \\
LOO & mcycle\_gp & GP regression & 133 & 66 & CV bridge \\
LOO & nabiximols (+int-qu) & beta-binomial & 257 & 114 & CV bridge \\
LOO & phase2a & Poisson, 4 param. & 262 & 4 & CV bridge + MM \\
LOO & phase3 & linear regression & 60 & 32 & CV bridge \\
LOO & roaches (+int-qu) & Poisson, (1\textbar obs) & 262 & 267 &
integrated PSIS \\
LOO & roaches\_nb & negative-binomial & 262 & 5 & CV bridge \\
LOO & roaches\_p & Poisson & 262 & 4 & CV bridge \\
LOO & sleep (+int-an) & normal, hier. & 144 & 42 & CV bridge \\
LOO & sleep\_t (+int-qu) & Student-\(t\), hier. & 144 & 43 & CV
bridge \\
LOO & student\_r2d2 & normal, R2D2 prior & 407 & 55 & CV bridge \\
LOO & voice & normal, horseshoe & 126 & 626 & CV bridge \\
LOGO & bangladesh\_grp (+int-srqis) & Bernoulli & 60 & 125 & CV
bridge \\
LOGO & gcse\_bb\_grp (+int-qu) & beta-binomial & 73 & 77 & CV bridge \\
LOGO & gcse\_n\_grp (+int-an) & normal & 73 & 152 & CV bridge + MM \\
LOGO & nabiximols\_grp (+int-qu) & beta-binomial & 105 & 114 & CV
bridge \\
LOGO & pupil\_grp (+int-an) & normal & 20 & 46 & bridge sampling
(int.) \\
LOGO & pupil\_sigma\_grp (+int-anqu) & normal, varying scale & 20 & 67 &
CV bridge (int.) \\
LOGO & sleep\_grp (+int-an) & normal & 18 & 42 & CV bridge + MM \\
LOGO & sleep\_t\_grp (+int-qu) & Student-\(t\) & 18 & 43 & CV bridge \\
LOGO & verbagg\_grp (+int-srqis) & ordered-logistic & 316 & 1601 & CV
bridge + MM \\
\end{longtable}

Table~\ref{tbl-int-cells} gives the integration rule of each integrated
LOGO cell (Section~\ref{sec-integration}). A moment-matched
Gauss--Hermite rule centers and scales its nodes by the mean and
covariance of the integrand, estimated in a pilot pass; a mode-centered
rule uses the mode and the curvature there.

\begin{longtable}[]{@{}
  >{\raggedright\arraybackslash}p{(\linewidth - 6\tabcolsep) * \real{0.2235}}
  >{\raggedright\arraybackslash}p{(\linewidth - 6\tabcolsep) * \real{0.2353}}
  >{\raggedleft\arraybackslash}p{(\linewidth - 6\tabcolsep) * \real{0.0588}}
  >{\raggedright\arraybackslash}p{(\linewidth - 6\tabcolsep) * \real{0.4824}}@{}}
\caption{Integration rule of each integrated LOGO cell. Dim. is the
dimension of the integral per group; 2+1 is an analytic two-dimensional
block and a one-dimensional
quadrature.}\label{tbl-int-cells}\tabularnewline
\toprule\noalign{}
\begin{minipage}[b]{\linewidth}\raggedright
Cell
\end{minipage} & \begin{minipage}[b]{\linewidth}\raggedright
Data model
\end{minipage} & \begin{minipage}[b]{\linewidth}\raggedleft
Dim.
\end{minipage} & \begin{minipage}[b]{\linewidth}\raggedright
Integration rule
\end{minipage} \\
\midrule\noalign{}
\endfirsthead
\toprule\noalign{}
\begin{minipage}[b]{\linewidth}\raggedright
Cell
\end{minipage} & \begin{minipage}[b]{\linewidth}\raggedright
Data model
\end{minipage} & \begin{minipage}[b]{\linewidth}\raggedleft
Dim.
\end{minipage} & \begin{minipage}[b]{\linewidth}\raggedright
Integration rule
\end{minipage} \\
\midrule\noalign{}
\endhead
\bottomrule\noalign{}
\endlastfoot
sleep\_grp, gcse\_n\_grp, pupil\_grp & normal & 2 & analytic \\
nabiximols\_grp & beta-binomial & 1 & moment-matched Gauss--Hermite, 80
nodes \\
gcse\_bb\_grp & beta-binomial & 1 & moment-matched Gauss--Hermite, 80
nodes \\
sleep\_t\_grp & Student-\(t\) & 2 & moment-matched tensor
Gauss--Hermite, 120 nodes \\
pupil\_sigma\_grp & normal, varying scale & 2+1 & analytic mean block
and mode-centered Gauss--Hermite, 15 nodes \\
bangladesh\_grp & Bernoulli & 2 & Laplace with importance sampling
correction \\
verbagg\_grp & ordered-logistic & 5 & Laplace with importance sampling
correction \\
\end{longtable}

The importance sampling correction of the Laplace approximation uses
\(L = 64\) points and \(J = 2\) randomizations per draw and group, that
is, 128 evaluations of the integrand plus a few to set the proposal.

\subsection{The self-diagnostic of every
method}\label{sec-methods-table}

Table~\ref{tbl-methods} collects, for every method of this paper, how it
computes each part of its self-diagnostic (Section~\ref{sec-framework}).

\begin{longtable}[]{@{}
  >{\raggedright\arraybackslash}p{(\linewidth - 10\tabcolsep) * \real{0.1548}}
  >{\raggedright\arraybackslash}p{(\linewidth - 10\tabcolsep) * \real{0.1310}}
  >{\raggedright\arraybackslash}p{(\linewidth - 10\tabcolsep) * \real{0.1310}}
  >{\raggedright\arraybackslash}p{(\linewidth - 10\tabcolsep) * \real{0.1310}}
  >{\raggedright\arraybackslash}p{(\linewidth - 10\tabcolsep) * \real{0.2262}}
  >{\raggedright\arraybackslash}p{(\linewidth - 10\tabcolsep) * \real{0.2262}}@{}}
\caption{The self-diagnostic of every method. The thresholds 0.7 for
Pareto-\(\hat{k}\) and 0.3 for the bridge MCSE follow earlier work; the
others are our choices. }\label{tbl-methods}\tabularnewline
\toprule\noalign{}
\begin{minipage}[b]{\linewidth}\raggedright
Method
\end{minipage} & \begin{minipage}[b]{\linewidth}\raggedright
Extra sampling
\end{minipage} & \begin{minipage}[b]{\linewidth}\raggedright
Total MCSE
\end{minipage} & \begin{minipage}[b]{\linewidth}\raggedright
Bias predictor
\end{minipage} & \begin{minipage}[b]{\linewidth}\raggedright
Reliability checks
\end{minipage} & \begin{minipage}[b]{\linewidth}\raggedright
Trusted when
\end{minipage} \\
\midrule\noalign{}
\endfirsthead
\toprule\noalign{}
\begin{minipage}[b]{\linewidth}\raggedright
Method
\end{minipage} & \begin{minipage}[b]{\linewidth}\raggedright
Extra sampling
\end{minipage} & \begin{minipage}[b]{\linewidth}\raggedright
Total MCSE
\end{minipage} & \begin{minipage}[b]{\linewidth}\raggedright
Bias predictor
\end{minipage} & \begin{minipage}[b]{\linewidth}\raggedright
Reliability checks
\end{minipage} & \begin{minipage}[b]{\linewidth}\raggedright
Trusted when
\end{minipage} \\
\midrule\noalign{}
\endhead
\bottomrule\noalign{}
\endlastfoot
Brute-force & one MCMC run per fold & truncGPD, RSS over folds &
downward & none & predicted RMSE \(<\) 1 nat \\
Brute-force + MM & as brute-force & truncGPD on matched folds & downward
& \(\hat{k}_{\mathrm{mm}} \leq 0.7\) to keep a matched fold & predicted
RMSE \(<\) 1 nat \\
CV bridge & as brute-force & IJ & none & finite, MCSE \(<\) 0.3, gap
\(\leq\) 50 nat (per fold) & every fold passes \\
PSIS & none & IJ & upward & \(\hat{k}_{\mathrm{psis}} \leq 0.7\) &
predicted RMSE \(<\) 1 nat and every
\(\hat{k}_{\mathrm{psis}} \leq 0.7\) \\
PSIS + MM & none & as PSIS & as PSIS (estimate of PSIS is reported) &
fewer than 8\% of folds with \(\hat{k}_{\mathrm{psis}} > 0.7\);
\(\hat{k}_w \leq 0.7\) & as PSIS, with warnings withdrawn \\
MixIS & one MCMC run on the mixture & IJ with common term & upward
(\(v\)) and downward (common) & \(\hat{k}(u)\) warns; MCMC convergence
incl.~\(\mathrm{ESS}(u)\) checked first & predicted RMSE \(<\) 1 nat \\
bMixIS & as MixIS & as MixIS & as MixIS & as MixIS & as MixIS \\
Integrated methods & none (integral after MCMC) & as the method & as the
method & as the method; the integral itself is not checked & as the
method \\
\end{longtable}

\subsection{\texorpdfstring{Choice of the largest plausible value
\(p_{\max}\)}{Choice of the largest plausible value p\_\{\textbackslash max\}}}\label{sec-pmax}

The truncGPD MCSE of brute-force needs a largest plausible value
\(p_{\max}\) of the values \(p(y_i \mid \theta_s)\)
(Section~\ref{sec-diag-bf}). We use the smaller of the physical or soft
bound and the full-data maximum
\(\max_s p(y_i \mid \theta^{\mathrm{f}}_s)\). This appendix shows why
the full-data maximum is needed.

The physical and soft bounds depend only on the data model. On folds
with a strong influence, they can be orders of magnitude larger than any
value the posterior reaches. This happens especially in LOGO. There the
bound is a product over the observations of the group, and the product
of the largest densities is much larger than the joint density of the
group under the posterior. Without the left-out data, the leave-fold-out
posterior usually moves away from \(y_i\), so its largest value of
\(p(y_i \mid \theta_s)\) usually stays below the full-data maximum.

\begin{figure}[t]

\centering{

\includegraphics[width=0.8\linewidth,height=\textheight,keepaspectratio]{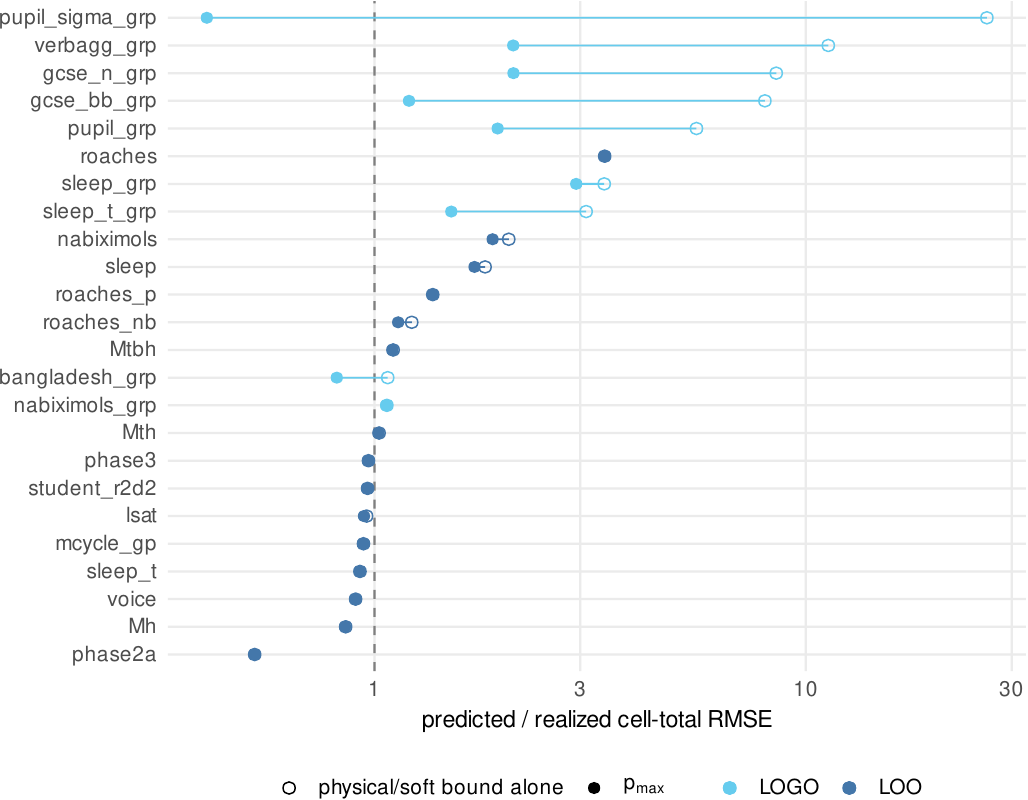}

}

\caption{\label{fig-pmax}Predicted against realized RMSE of brute-force
with the physical or soft bound alone (open circles) and with
\(p_{\max}\) (filled circles), per cell (\(S = 4000\), 100 runs); log
scale; dashed line: 1, calibrated; right of it, the error is
overpredicted.}

\end{figure}

Figure~\ref{fig-pmax} compares the two choices on every cell with a
brute-force result. On the LOGO cells gcse\_n\_grp, gcse\_bb\_grp,
sleep\_grp and sleep\_t\_grp, the realized error is small. The physical
or soft bound alone overpredicts the RMSE there 3.1 to 8.5 times, and
the trust rule would reject accurate runs. With the full-data maximum,
the overprediction falls to 1.2 to 2.9 times, but gcse\_n\_grp and
sleep\_grp are still rejected (Section~\ref{sec-limitations}). On the
cells with a large error, both choices reject. On verbagg\_grp and
pupil\_grp, \(p_{\max}\) is closer to the realized RMSE: the
overprediction falls from 11 to 2.1 times and from 5.6 to 1.9 times. On
pupil\_sigma\_grp, the physical or soft bound overpredicts 26 times, and
\(p_{\max}\) underpredicts 2.4 times, but its prediction of 21 nat is
still far above 1 nat. On most LOO cells, the two choices give the same
prediction. On phase2a, both choices give too small a prediction,
because the full-data posterior puts almost no mass where the left-out
observation is likely.

\subsection{Moment matching for brute-force}\label{sec-bf-mm}

We apply moment matching to brute-force as defined in
Section~\ref{sec-framework-mm}: only on folds where
\(|\mathrm{bias}_{\mathrm{pred},i}| > 1\) nat or
\(\hat{k}_{\mathrm{bf}} > 1\), and a matched estimate is kept only when
\(\hat{k}_{\mathrm{mm}} \leq 0.7\).

On roaches\_p, the error changes from \(-17.1\) to \(-0.55\) nat, and
95\% of runs are trusted. On phase2a, it changes from \(-78.0\) to
\(-0.75\) nat, and 63\% of runs are trusted, although the realized RMSE
is 1.08 nat. On the other cells, moment matching changes the estimate by
at most 0.02 nat. On the pupil cells, \(\hat{k}_{\mathrm{mm}}\) is
always above 0.7, so no matched estimate is kept and moment matching
cannot help. A check refitted 81 folds where a matched estimate was kept
(on roaches\_p, phase2a, gcse\_bb\_grp and sleep\_t\_grp). On every one
of them, the accepted transform was the simplest one, a shift of the
mean.

The two rules are essential. On a fold with an extreme tail, the shift
can move the draws into a region where the leave-fold-out posterior has
little mass, and the matched estimate then has a large downward bias, in
the same direction as brute force. Applied to the roaches folds with
\(\hat{k}_{\mathrm{bf}} > 0.7\) and always kept, moment matching changes
the error from \(-0.65\) to \(-12.67\) nat.

\subsection{Covariance between folds in the total
MCSE}\label{sec-ij-rss}

\begin{figure}[H]

\centering{

\includegraphics[width=1\linewidth,height=\textheight,keepaspectratio]{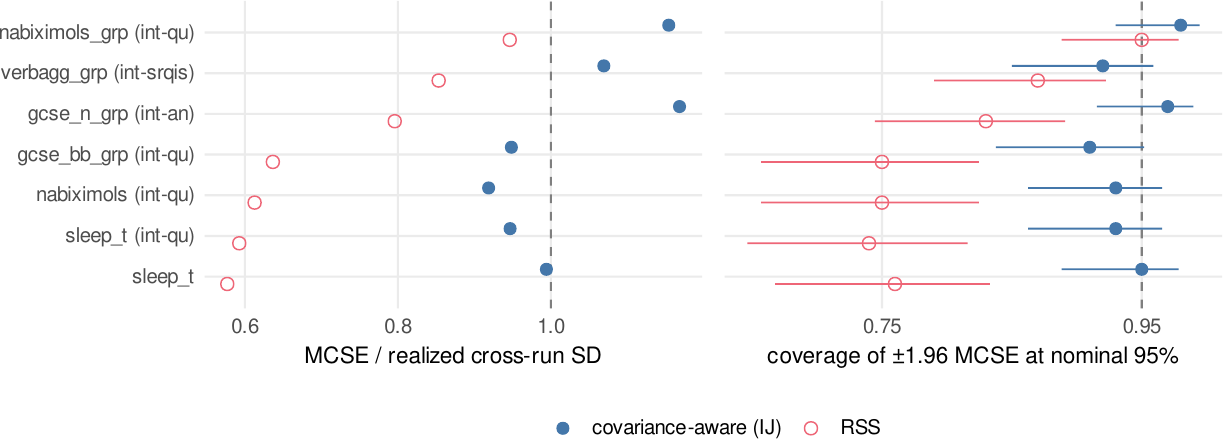}

}

\caption{\label{fig-ij-rss}The IJ total MCSE matches the realized SD;
the RSS total is too small. IJ and RSS total MCSE against the realized
SD (left) and coverage of the symmetric interval (right), on the seven
PSIS-based combinations with more than 90\% of runs trusted; dashed
lines: 1 (left) and 0.95 (right).}

\end{figure}

All folds of PSIS use the same draws, so their errors are correlated.
Figure~\ref{fig-ij-rss} compares two total MCSEs on the seven
combinations of a PSIS-based method and a cell with more than 90\% of
runs trusted. The RSS (open red circles) adds only the fold variances.
The IJ total (filled blue points) also adds the covariances. The left
panel shows the ratio of the total MCSE to the realized SD, where 1 is
exactly right. The IJ total lies at 0.92--1.17, the RSS at only
0.58--0.95. With the RSS, a 95\% interval contains the error in only
81\% of the runs, against 94\% with the IJ total (right panel).

\end{document}